\pdfoutput=1

\documentclass[11pt]{article}

\usepackage[preprint]{acl}

\usepackage{algorithm}
\usepackage{algpseudocode}
\usepackage{amsmath}
\usepackage{booktabs}

\usepackage{amssymb}    
\usepackage{amsfonts}   
\usepackage{amsthm}

\newtheorem{definition}{Definition}
\newtheorem{property}{Property}

\usepackage{subcaption}
\usepackage{CJKutf8}

\usepackage{color}

\usepackage{times}
\usepackage{latexsym}

\definecolor{xhsbg}{RGB}{240, 248, 255}  
\definecolor{xhsred}{RGB}{220, 20, 60}   
\usepackage{tcolorbox}
\tcbuselibrary{skins, breakable}
\usepackage{caption} 

\usepackage{longtable}
\usepackage{tabularx}
\usepackage{multirow}
\usepackage{array}
\usepackage{longtable}
\usepackage{ragged2e}
\usepackage{listings}
\usepackage{makecell}

\newcolumntype{Y}{>{\RaggedRight\arraybackslash}X}
\newtcolorbox{ACLPrompt}[1][]{
    enhanced,
    breakable,         
    width=\linewidth,  
    colback=gray!5!white, 
    colframe=black!75,    
    title=Prompt,
    boxrule=0.8pt,        
    arc=2pt,              
    left=4pt, right=4pt, top=4pt, bottom=4pt, 
    #1
}

\newtcolorbox{finding}[1]{
  before={\par\noindent},
  after={\par\noindent},
  colback=xhsbg!10,
  colframe=xhsred!70,
  title=Finding #1,
  fonttitle=\bfseries
}

\usepackage[T1]{fontenc}

\usepackage[utf8]{inputenc}

\usepackage{microtype}

\usepackage{inconsolata}

\usepackage{graphicx}

\title{Scaling Behavior of Single LLM-Driven Multi-Agent Systems}

\author{
  Jialing Li \\
  Fudan University \\
  \texttt{jialingli22@m.fudan.edu.cn}
  \And
  Zhouhong Gu \\
  Fudan University \\
  \texttt{zhgu22@m.fudan.edu.cn}
  \And
  Yin Cai \\
  Fudan University \\
  \texttt{ycai25@m.fudan.edu.cn}
  \AND
  Hongwei Feng\thanks{Corresponding author} \\
  Fudan University \\
  \texttt{hwfeng@fudan.edu.cn}
}

\begin{document}
\begin{CJK}{UTF8}{gbsn}
\maketitle

\begin{abstract}

The burgeoning field of LLM-based Multi-Agent Systems (MAS) promises to tackle complex tasks through collaborative intelligence, yet fundamental questions regarding their scaling behavior and intrinsic collective dynamics remain underexplored. This paper systematically investigates how the performance of a homogeneous MAS evolves as the number of agents increases, isolating the variable of collaboration from model or knowledge heterogeneity. We propose the Sequential Iterative Multi-Agent System (SIMAS) framework, a minimalist architecture centered on sequential inter-agent communication, to clearly observe scaling effects. Through extensive experiments across diverse tasks and model scales, we establish that MAS performance does not scale monotonically with agent count but follows a pattern of diminishing returns, governed by a trade-off between collaborative synergy and coordination overhead. Our findings reveal that effective MAS requires a sufficiently capable base LLM, that task type critically modulates the optimal agent count, and that collective intelligence is an emergent property contingent on strategic interaction design rather than a guaranteed outcome of agent plurality. The performance degradation stems coordination overhead rather than merely long-context failure, and the scaling tendency generalizes across interaction architectures like structured debate topologies. This work provides a foundational understanding of MAS scaling laws, offering practical guidance for designing efficient collaborative systems and challenging the prevailing assumption that more agents invariably lead to better performance.

\end{abstract}

\section{Introduction}
\label{sec:intro}
In recent years, Large Language Models (LLMs) have demonstrated remarkable capabilities in text generation, complex reasoning, and decision-making, establishing themselves as the core foundation for constructing intelligent systems, often referred to as "agents".
Though individual agent has exhibited excellent problem-solving ability in expansive fields with performance enhanced—for instance, through Chain-of-Thought (CoT) prompting \cite{wei2022chain} to elicit step-by-step reasoning, or by enabling models to leverage external APIs via frameworks like Toolformer \cite{schick2023toolformer}, however, many real-world challenges, such as sophisticated software development or multi-faceted problem-solving, inherently require collaborative efforts.
This necessity has driven the emergence of LLM-based Multi-Agent Systems (MAS), a field that has rapidly evolved from early exploratory frameworks to complex systems \cite{xi2023survey, luo2025survey}, where multiple agents interact to achieve common goals.

\begin{figure*}[t]
    \centering
    \includegraphics[width=\textwidth]{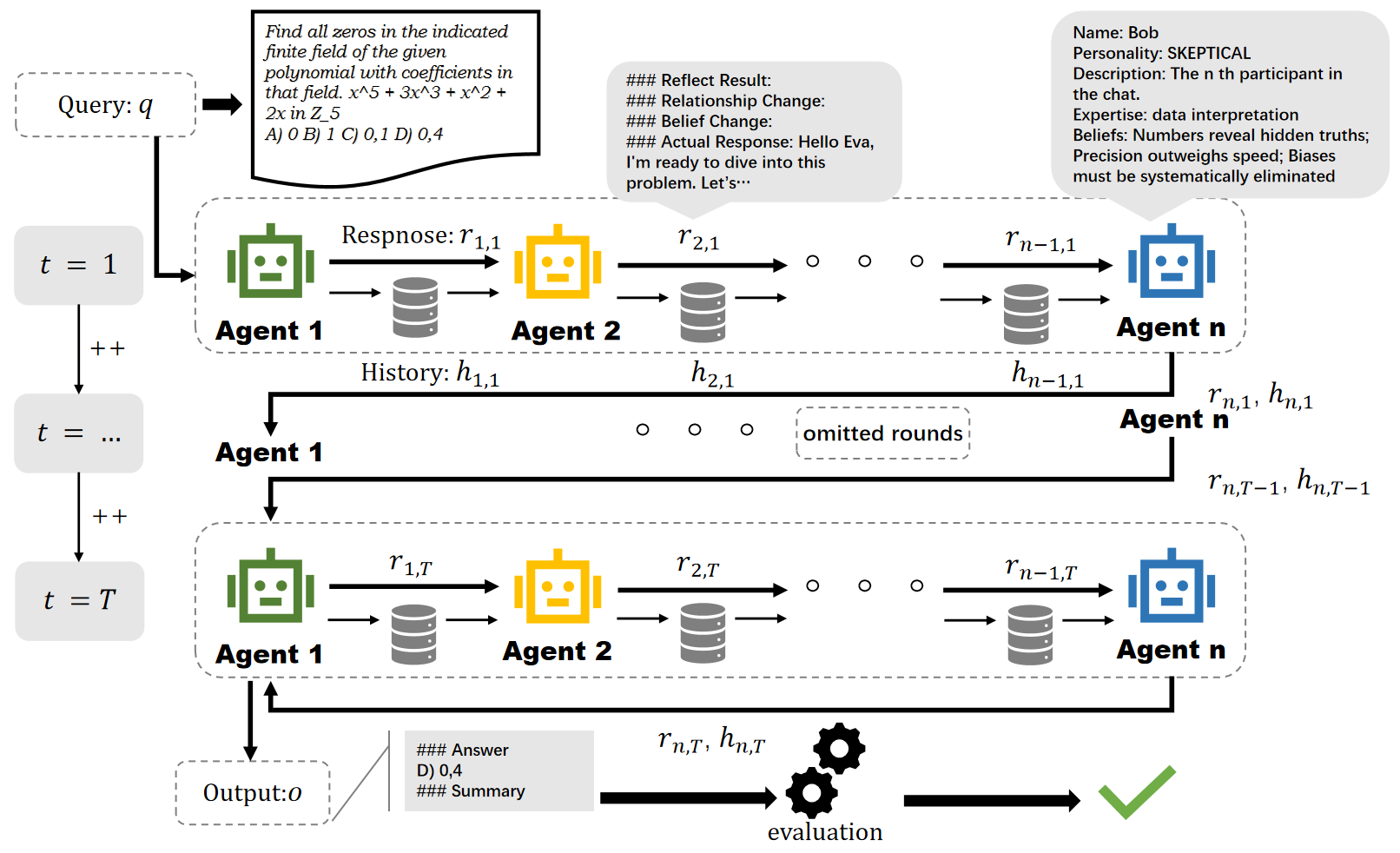}
    \caption{Workflow of the SIMAS. A group of $n$ agents ($a_1, a_2, ..., a_n$), each configured with a distinct profile (personality, core beliefs, expertise), engages in $T$ rounds of sequential discussion. In each round, every agent generates a response based on the progressively accumulated conversation history $h_{i-1,t}$, which monotonically expands to $h_{n,t}$. After $T$ rounds, the first agent $a_1$ synthesizes the final output $o$ from the complete history $h_{n,T}$.}
    \label{fig:multi-agent-system}
\end{figure*}

In MAS, collaboration enables agents to share knowledge, assign specialized roles, and integrate their complementary strengths, thereby tackling tasks beyond the scope of any single agent—ranging from handling extensive contexts \cite{zhang2025chain} to complex social simulations \cite{qian2023communicative}.
This vision is supported by a growing series of frameworks:
AutoGen \cite{autogen} facilitates customizable agent conversations, CAMEL \cite{camel} explores role-playing for idea exploration, and MetaGPT \cite{hong2023metagpt} assigns standardized workflows to simulate a software company.
These systems often deploy agents as "reasoning experts" or "fact-checkers" to boost collective output.
However, despite these engineering advances and recent efforts to probe the robustness of agent communication \cite{he2025red}, a fundamental scientific understanding of MAS collective behavior remains under-explored.
\textit{First}, in order to isolate the scaling effects of collaboration without the confounding variable of model heterogeneity, we ask: when all agents are built using the same underlying language model, does increasing the number of agents consistently improve the system's performance?
\textit{Second}, are there common patterns in how these systems operate, regardless of how they are designed?
\textit{Third}, how do the collaboration of MAS differ from the reasoning of a single agent?

To provide a systematic analysis of the scaling laws in collaboration of MAS, we establish a investigation into how the performance of MAS evolves as the number of agents increases.
We propose the Sequential Iterative Multi-Agent System (SIMAS) framework, which strips away scenario-specific components and retains only sequential inter-agent communication as its core mechanism.
This design minimizes architectural complexity, enabling clear observation of performance scaling with agent count and identification of general patterns in MAS.
Through a rigorous comparison between MAS and single-agent baselines, we aim to quantitatively delineate the boundaries where collaboration yields positive returns versus where it introduces diminishing returns.

Our investigation yields several principal findings:
\textit{Model capability is a prerequisite for effective MAS.}
We observe that only models above a certain performance threshold (typically larger-scale models) can effectively power MAS.
\textit{LLM-based MAS scalability is governed by a fundamental trade-off between collaborative synergy and coordination overhead.}
This reframes the goal from "more agents" to "optimal number," highlighting that excessive collaboration can be detrimental, particularly in reasoning-intensive tasks where focus and coherence are critical; performance degradations derive from coordination overhead rather than merely long-context failure, and the tendency generalizes across interaction architectures like structured debate topologies.
\textit{Collective intelligence in MAS is an emergent property contingent on interaction architecture, not an automatic outcome of agent plurality.}
Without strategic architectural design, an MAS risks achieving only the illusion of collaboration while failing to surpass the capabilities of a well-prompted individual.

\section{Related Works}

\subsection{Architectures for Multi-Agent Collaboration}
The foundation of MAS lies in defining how agents interact to extend the capabilities of individual LLM.
Early research focused on unstructured communicative frameworks.
For instance, CAMEL \cite{camel} introduced "inception prompting" to facilitate autonomous role-playing, demonstrating that distinct personas can guide solution exploration.
Building on this, AutoGen \cite{autogen} provided a flexible infrastructure for dynamic conversations between agents, while MetaGPT \cite{hong2023metagpt} incorporated Standard Operating Procedures (SOPs) with rigid role assignments to streamline complex workflows like software generation.
Similarly, AgentVerse \cite{chen2023agentverse} and the recent Agent-Pro \cite{zhang2024agent} introduce iterative optimization mechanisms, where agents engage in "reflect-and-refine" loops to correct errors dynamically.


Despite the diversity of existing multi-agent architectures, they all fundamentally rely on a core collaborative process involving iterative communication, shared contextual memory, and sequential reasoning \cite{xi2025rise, xi2023survey, guo2024large}. Advanced mechanisms such as dynamic reflection are merely engineering scaffolds to mitigate the inherent friction of this core, meaning all such systems belong to the same interactive paradigm \cite{talebirad2023multi}. By designing SIMAS as a minimalist system that isolates this core \cite{xi2025rise}, we reveal that the trade-off between collaborative synergy and coordination overhead is a universal property of all multi-agent systems, challenging the assumption that adding more agents always improves performance and underscoring the importance of architectural design.

\subsection{Scaling Behavior in Multi-Agent Systems}
Recently, the community has begun to formally investigate the scaling laws and inherent limitations of LLM-based Multi-Agent Systems (MAS). 
A notable concurrent work~\cite{qian2025scaling} explores scaling multi-agent collaboration by structuring agents into a Directed Acyclic Graph (DAG) topology, termed MacNet. 
By enforcing localized information passing they successfully mitigate context explosion and observe a logistic growth pattern as the system scales to hundreds or thousands of agents. 

Conversely, other recent studies have begun to critically examine the inherent vulnerabilities of MAS as they scale. 
~\cite{flint2025groupsize} investigate group size effects, revealing that expanding agent counts can paradoxically exacerbate collective misalignment and consensus failure. 
Similarly, ~\cite{cemri2025why} provide a comprehensive taxonomy of why MAS fail, highlighting critical issues such as reasoning fragmentation, compounding hallucinations, and cascading errors during complex interactions.
Complementary to the coordination overhead perspective, a concurrent work by \cite{shehata2026bystander} reveals an orthogonal failure mode: the algorithmic ``Bystander Effect'' in MAS, where simulated social pressure triggers cognitive loafing: agents abandon independent reasoning to sycophantically comply with swarm consensus even when internally deriving correct answers. 
Together, these works establish that multi-agent collaboration faces both \textit{coordination} and \textit{social-psychological} challenges, with effective MAS requiring architectural safeguards against both.

While these studies offer crucial insights, they leave a critical gap in understanding the raw, unmitigated scaling dynamics of MAS. 
Our work bridges this gap by introducing the SIMAS framework as a deliberate, minimalist control group. By intentionally stripping away complex topologies, information-filtering mechanisms, and structured aggregation, SIMAS utilizes pure sequential communication with progressive history accumulation to empirically quantify the tension collaborative synergy and coordination overhead to demonstrate the scaling behavior of unstructured multi-agent dialogue.

\section{Methodology}
\subsection{Preliminaries}
To provide a clear overview of our research question, we first define the notations and the mathematical formulation of the Multi-Agent System (MAS). The key symbols used throughout this paper are listed in Table~\ref{tab:notation}.

\begin{table}[t]
\centering
\resizebox{\columnwidth}{!}{
\begin{tabular}{ll}
\toprule
\textbf{Symbol} & \textbf{Definition} \\
\midrule
$\mathcal{Q}$ & Input question space \\
$q$ & A specific input question instance ($q \in \mathcal{Q}$) \\
$\mathcal{A}$ & The set of agents $\{a_1, a_2, \dots, a_n\}$ \\
$n$ & Total number of agents \\
$T$ & Total number of discussion rounds \\
$\mathcal{H}$ & Conversation history space \\
$h_{i,t}$ & Specific conversation history after agent $a_i$ in round $t$ \\
$\Pi$ & Response generation function (LLM) \\
$\mathcal{O}$ & Output space \\
$o$ & Final answer instance ($o \in \mathcal{O}$) \\
\bottomrule
\end{tabular}}
\caption{Notations used in the SIMAS framework.}
\label{tab:notation}
\end{table}

\begin{definition}[Multi-Agent System Formulation]
The multi-agent system $\mathcal{M}$ is formally defined as a sextuple:
\[
\mathcal{M} = \langle \mathcal{A}, \mathcal{Q}, \mathcal{H}, T, \Pi, \mathcal{O} \rangle
\]
where $\mathcal{A} = \{a_1, a_2, \dots, a_n\}$ denotes the set of agents, with $n \in \mathbb{N}^+$ representing the number of agents;
$\mathcal{Q}$ denotes the input question space;
$\mathcal{H}$ denotes the conversation history space;
$T \in \mathbb{N}^+$ denotes the number of discussion rounds;
$\Pi: \mathcal{H} \times \mathcal{A} \rightarrow \mathcal{R}$ denotes the response generation function, where $\mathcal{R}$ is the response space;
and $\mathcal{O}: \mathcal{H} \rightarrow \mathcal{A}$ denotes the final answer synthesis function.
\end{definition}

\subsection{The SIMAS Framework}
Based on the formulation above, we designed a specific implementation named Sequential Iterative Multi-Agent-System (SIMAS).
This framework deliberately preserves only sequential inter-agent communication as a primary and minimalist baseline for MAS.
The process is structured as follows:
The discussion is conducted over $T$ rounds (default 3 rounds).
In each round, every agent generates a response based on the original question and the entire conversation history from all previous agents and rounds.

\begin{property}[Sequential Communication]
The system employs sequential inter-agent communication within each round:
\[
\forall t \in [1,T], \quad h_{n, t} = h_{1, t} \cup \bigcup_{i=2}^{n} \{\Pi(h_{i-1, t}, a_i)\}
\]
\end{property}

\begin{property}[Progressive History Accumulation]
The conversation history grows monotonically in one round:
\[
\forall t \in [1,T], h_{1,t} \subset h_{2,t} \subset \cdots \subset h_{n,t}
\]
\end{property}

After all agents have responded in a round, the process repeats for the next round, ensuring each agent has multiple opportunities to contribute to the discussion.
The final answer is synthesized by the first agent based on the complete conversation history.
The whole process of our MAS can be seen in Alg.~\ref{alg:collaboration_formal} and Figure~\ref{fig:multi-agent-system}.
Formal definitions of system output and agent configuration are provided in Appendix~\ref{app:formal_defs}.

\begin{algorithm}[t]
\caption{Multi-Agent Collaborative Process}
\label{alg:collaboration_formal}
\begin{algorithmic}[1]
\Require Question $q \in \mathcal{Q}$, agent set $\mathcal{A}$, rounds $T$
\Ensure Final answer $o$
\State Initialize history $h_0 \gets \{q\}$
\For{round $t = 1$ to $T$}
    \For{each agent $a_i \in \mathcal{A}$}
        \State $r_{i,t} \gets \Pi(h_{i-1, t}, a_i)$ \Comment{Generate response based on current history}
        \State $h_{i, t} \gets h_{i-1, t} \cup \{r_{i,t}\}$ \Comment{Update conversation history}
    \EndFor
\EndFor
\State \Return $\Pi(h_{n, T}, a_1)$ \Comment{First agent generates final answer based on complete history}
\end{algorithmic}
\end{algorithm}

\subsection{Setting of Single Agent}
In this experiment, each agent $a_i \in \mathcal{A}$ was driven by LLM and assigned descriptions, personalities, beliefs, and strengths, with the descriptions set as "the i-th assistent in the group chat" and therefore not considered as influencing factors.
All agents are powered by the same LLM, sharing identical model parameters and knowledge base.
To introduce diversity in reasoning styles, each agent is assigned a unique profile before the discussion begins. 

The profile consists of three attributes: \textbf{\textit{personality, core beliefs, and expertise}}. \textbf{\textit{Personality}} is a general disposition (e.g., "Skeptical" or "Aggressive") that guides the tone and focus of their reasoning, \textbf{\textit{expertise}} is a specific domain of knowledge (e.g., "Data Interpretation" or "Logical Reasoning") they are instructed to prioritize, and \textbf{\textit{core beliefs}} are a set of principles (e.g., "Precision outweighs speed") that influence their approach to problem-solving.

These attributes are automatically generated by the language model for each agent to simulate a group of diverse experts. The specific configurations and formal definitions of agent attributes are detailed in Appendix~\ref{app:formal_defs}.

\section{Experiments}
\begin{figure}[t]
    \centering
    \includegraphics[width=\columnwidth]{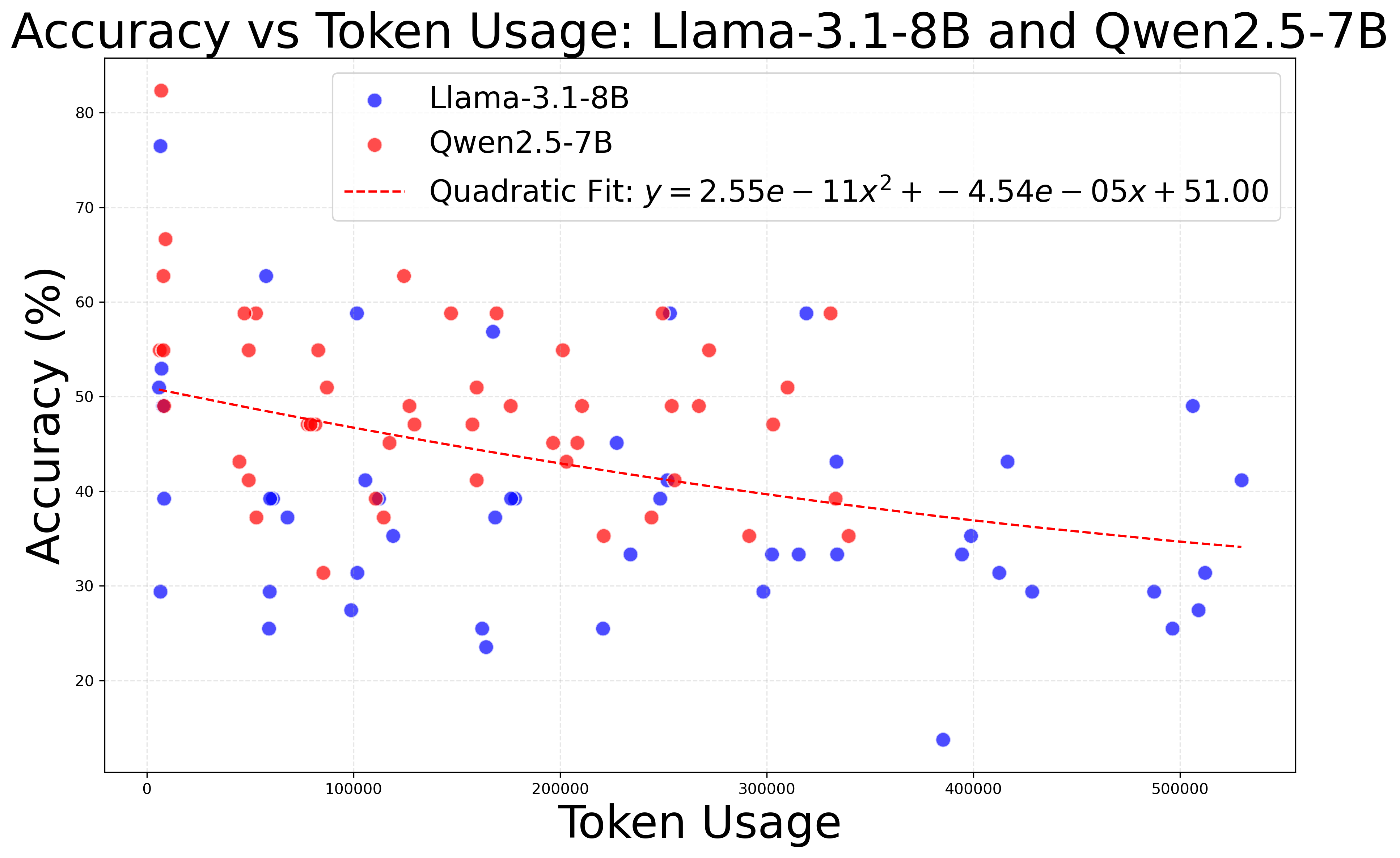}
    \caption{Result of models with small parameters on all subjects}
    \label{fig:model_scale1}
\end{figure}

\begin{figure}[t]
    \centering
    \includegraphics[width=\columnwidth]{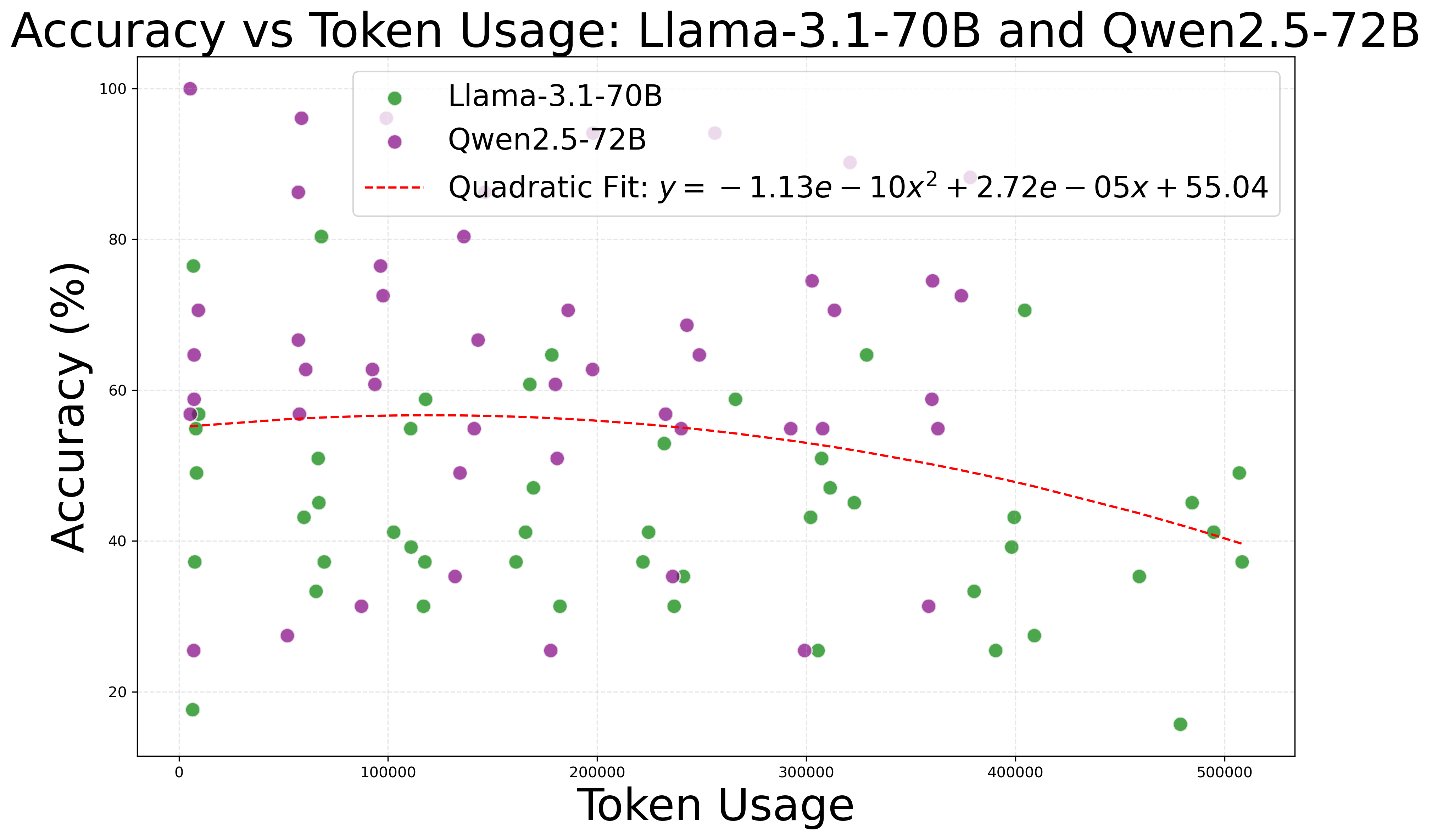}
    \caption{Result of models with large parameters on all subjects}
    \label{fig:model_scale2}
\end{figure}

\begin{figure*}[t]
    \centering
    \begin{subfigure}[b]{0.3\textwidth}
        \centering
        \includegraphics[width=\textwidth]{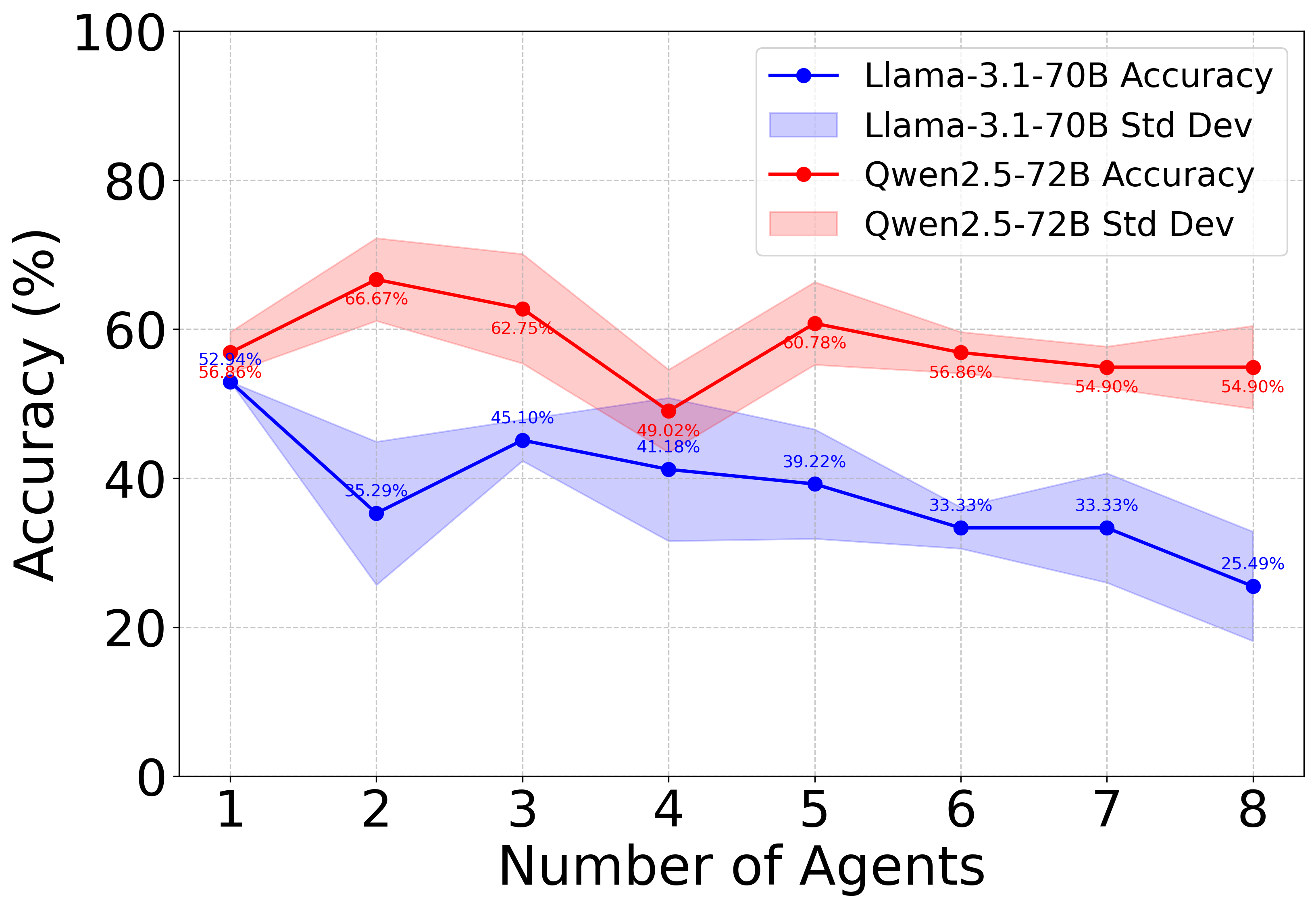}
        \caption{Performance on Abstract Algebra}
        \label{fig:task_type1}
    \end{subfigure}
    \hfill
    \begin{subfigure}[b]{0.3\textwidth}
        \centering
        \includegraphics[width=\textwidth]{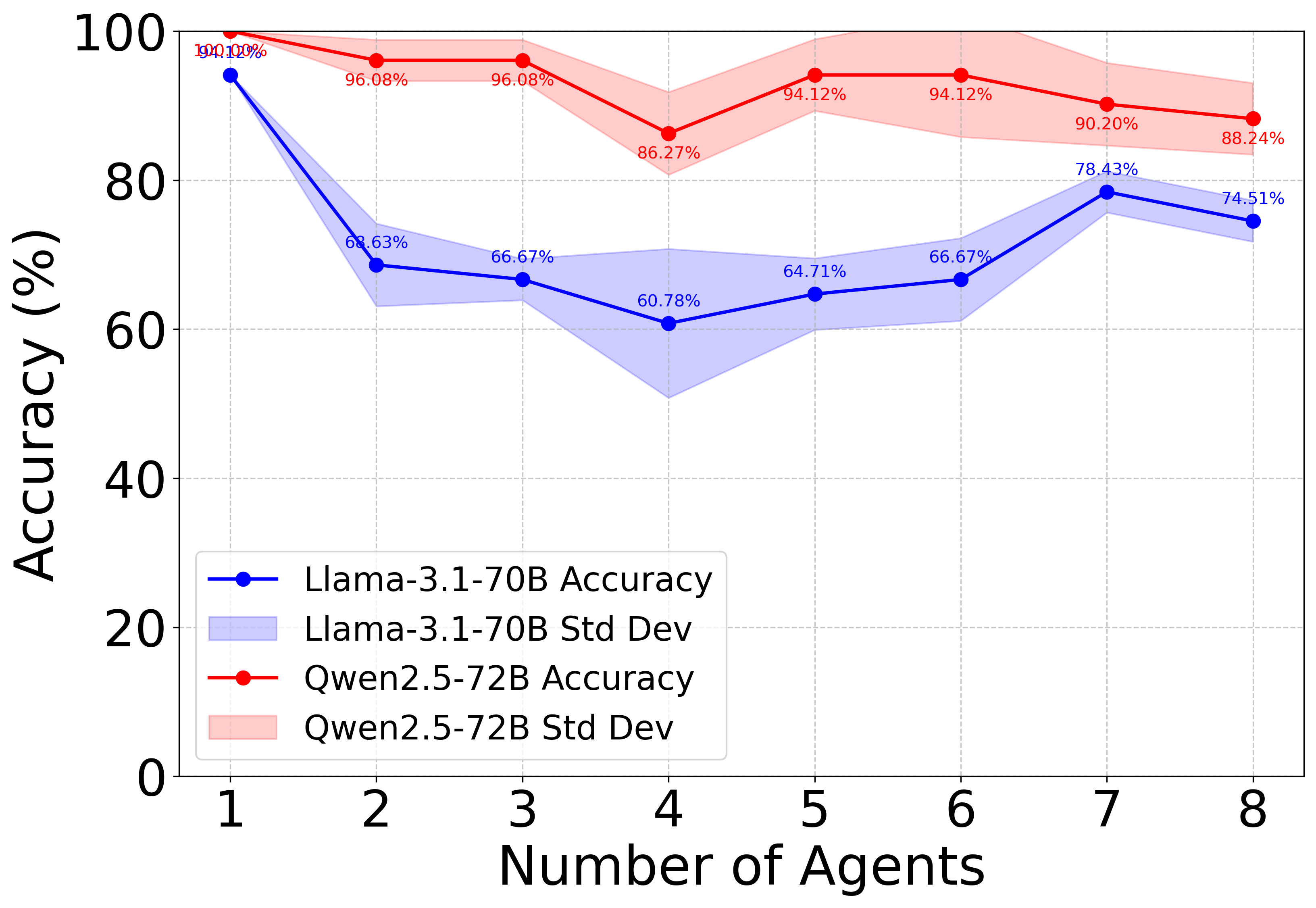}
        \caption{Performance on Philosophy}
        \label{fig:task_type2}
    \end{subfigure}
    \hfill
    \begin{subfigure}[b]{0.3\textwidth}
        \centering
        \includegraphics[width=\textwidth]{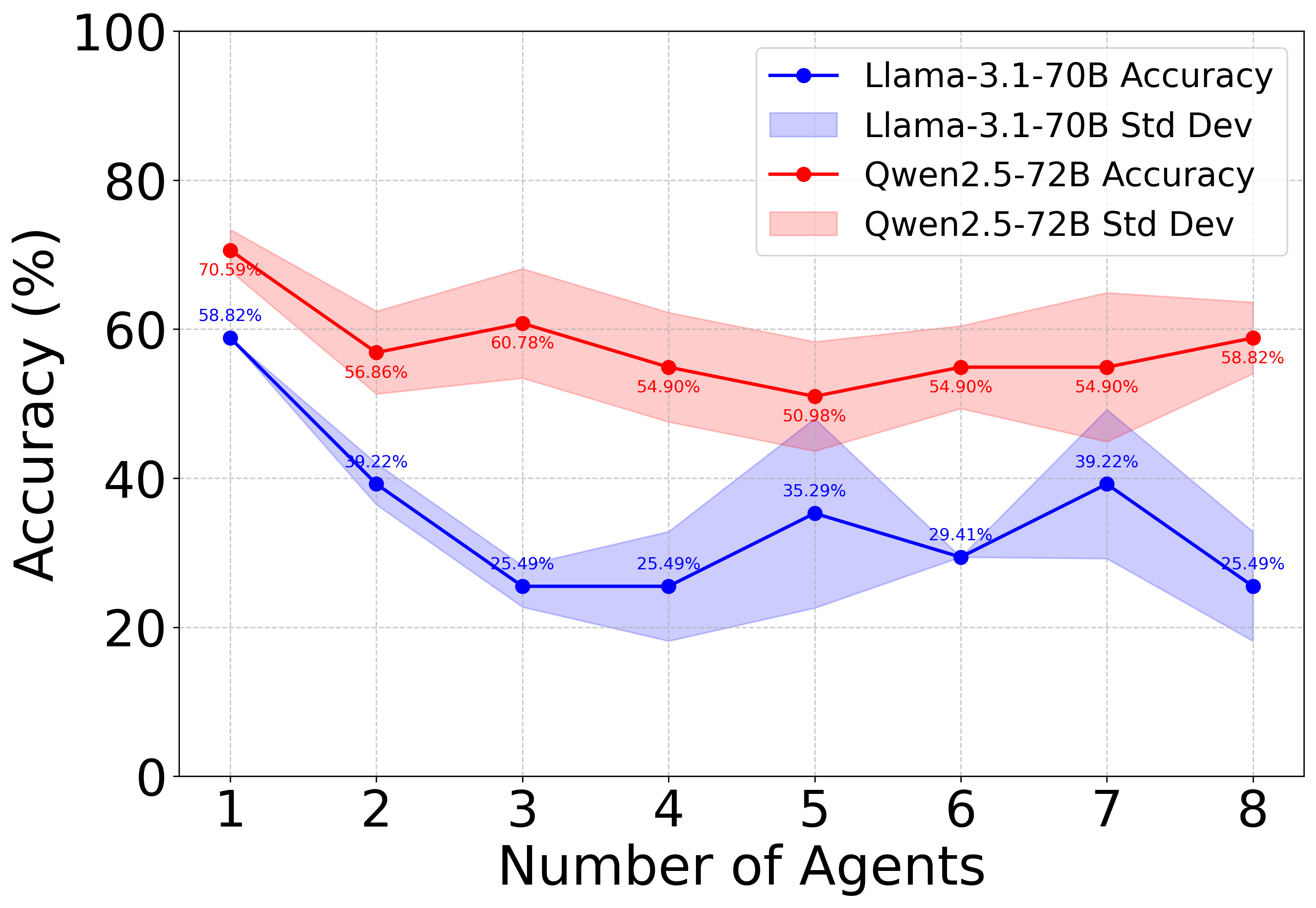}
        \caption{Performance on Formal Logic}
        \label{fig:task_type3}
    \end{subfigure}
    \caption{Task-type modulation of scaling on Llama-3.1-70B and Qwen2.5-72B. Reasoning tasks suffer sharp declines post-peak; knowledge tasks show more tolerance.}
    \label{fig:all_tasks}
\end{figure*}

\subsection{Experimental Setup}

To systematically examine the scaling effects of agent collaboration, we vary the number of agents $n$ from 1 to 8 in increments of 1.
This granular range allows us to capture the nuanced relationship between team size and performance, and it directly aligns with the practical band of deployed MAS like AutoGen, where the synergy–overhead trade-off is most critical.
Regarding agent configurations, we established a baseline condition where agents possess personality traits, expertise, and beliefs.

For performance evaluation, we employ three complementary benchmarks, each targeting distinct cognitive capabilities:
\textbf{The Massive Multitask Language Understanding (MMLU) dataset}~\cite{hendrycks2021measuring}, covering 57 diverse subjects across STEM, humanities, and social sciences, to assess general knowledge and broad reasoning capabilities.
\textbf{The E-KAR dataset}~\cite{luo2024ekar}, specifically designed for evaluating analogical reasoning through entity-based knowledge graph analogies.
\textbf{The AIME 2025}~\cite{aime2025} competition problems, consisting of advanced mathematical reasoning challenges, which we use in Section~5 to compare MAS against single-agent reasoning on highly complex tasks.

Note that in Section~5 for external literature comparison, we will exclusively reference additional datasets such as \textbf{GSM8K}, \textbf{MATH}, and \textbf{AIME 2024} and the data is directly cited from \cite{gu2025agentgroupchatv2} rather than our own primary experiments.

From MMLU, we select 6 representative disciplines (see Table~\ref{tab:subject_compare}) and sample 17 questions each to balance efficiency with robustness.
Each experimental configuration is repeated three times, with results averaged to ensure stability.

We employ two widely-used, open-source model families: Llama-3.1~\cite{llama3.1} and Qwen2.5~\cite{qwen2.5}.
To understand the impact of model scale, we use versions with 8B/7B and 70B/72B parameters from each family, and due to computational cost and inference speed constraints, we did not evaluate larger models beyond 70B parameters.
These models provide a strong baseline for general-purpose reasoning, allowing for a clear analysis of multi-agent scaling effects with balance of capabilities and generation speed.

To isolate the contribution of each attribute, we compare this baseline against four alternative configurations: one lacking all three attributes, and three configurations each removing exactly one of the three attributes.
This structured ablation enables us to analyze how each factor contributes to system performance and stability.
For these ablation studies, we specifically use the Llama-3.1-70B model with agent number $n \in \{2,4,6,8\}$, chosen to represent key points in the scaling range while reducing experimental overhead.

To address the potential confound between agent count and context length, we conduct a controlled token-padding experiment (Section~\ref{sec:context_control}) where total input length is fixed across all agent counts using neutral filler tokens.
To assess cross-architectural generality, we implement a parallel AutoGen GroupChat debate baseline (Section~\ref{sec:autogen}) with the same model--dataset combinations.

\subsection{Experimental Results and Analysis}

To investigate the scaling behavior of MAS performance $A(\mathcal{M}_n)$ with respect to agent count $n$, we first establish a criterion to distinguish productive collaboration from mere agent plurality.
\begin{definition}[sufficiently driven]
For a given task $T$ with evaluation metric $A: \mathcal{M} \rightarrow [0,1]$ measuring accuracy, let $A(\mathcal{M}_n)$ denote the accuracy of the system with $n$ agents.

Given a set of experimental conditions $C$ held constant, $\mathcal{M}$ is said to be \emph{sufficiently driven} if and only if:
\[
\exists\; n_1, n_2 \in \mathbb{N},\; n_1 \neq n_2,\; n_1, n_2 \geq 2 \quad
\quad
\]
\[
A(\mathcal{M}_{n_1}) \geq A(\mathcal{M}_1) \quad \text{and}
\quad A(\mathcal{M}_{n_2}) \geq A(\mathcal{M}_1)
\]

where the comparison is performed under identical conditions $C$.
\end{definition}
The rationale is twofold: the first point requires that the global peak is achieved by a multi-agent configuration ($n \ge 2$); the second precludes a single spurious spike from statistical noise.
A system failing this condition exhibits monotonic degradation, which is intrinsically linked to base model capability—small models (7B/8B) consistently fail (Figure~\ref{fig:model_scale1}).

\begin{finding}{1}
\textbf{If a MAS is sufficiently driven, the performance will scale with diminishing returns, following an inverted-U relationship with agent count.}
\end{finding}

For sufficiently driven MAS, performance initially improves due to collaborative synergy—diverse perspectives and mutual error correction—but peaks at an optimal $n^*$.
Beyond this point, coordination overhead (information redundancy, conflicting reasoning paths) dominates and causes decline.
This overhead is quantifiable: within SIMAS, token cost scales quadratically with $n$ (Figure~\ref{fig:model_scale2}).
The location of $n^*$ and decline steepness are modulated by several factors analyzed below.

To illustrate the impact of model scale, we compared the experimental results of large-parameter models and small-parameter models across all disciplines, with a particular focus on the representative case of College Physics for detailed analysis.
Results are demonstrated in Figure~\ref{fig:model_scale1} and~\ref{fig:model_scale2}.
More detailed information is shown in the Appendix~\ref{app:model_scale}.

\begin{figure}[t]
        \centering
        \includegraphics[width=\columnwidth]{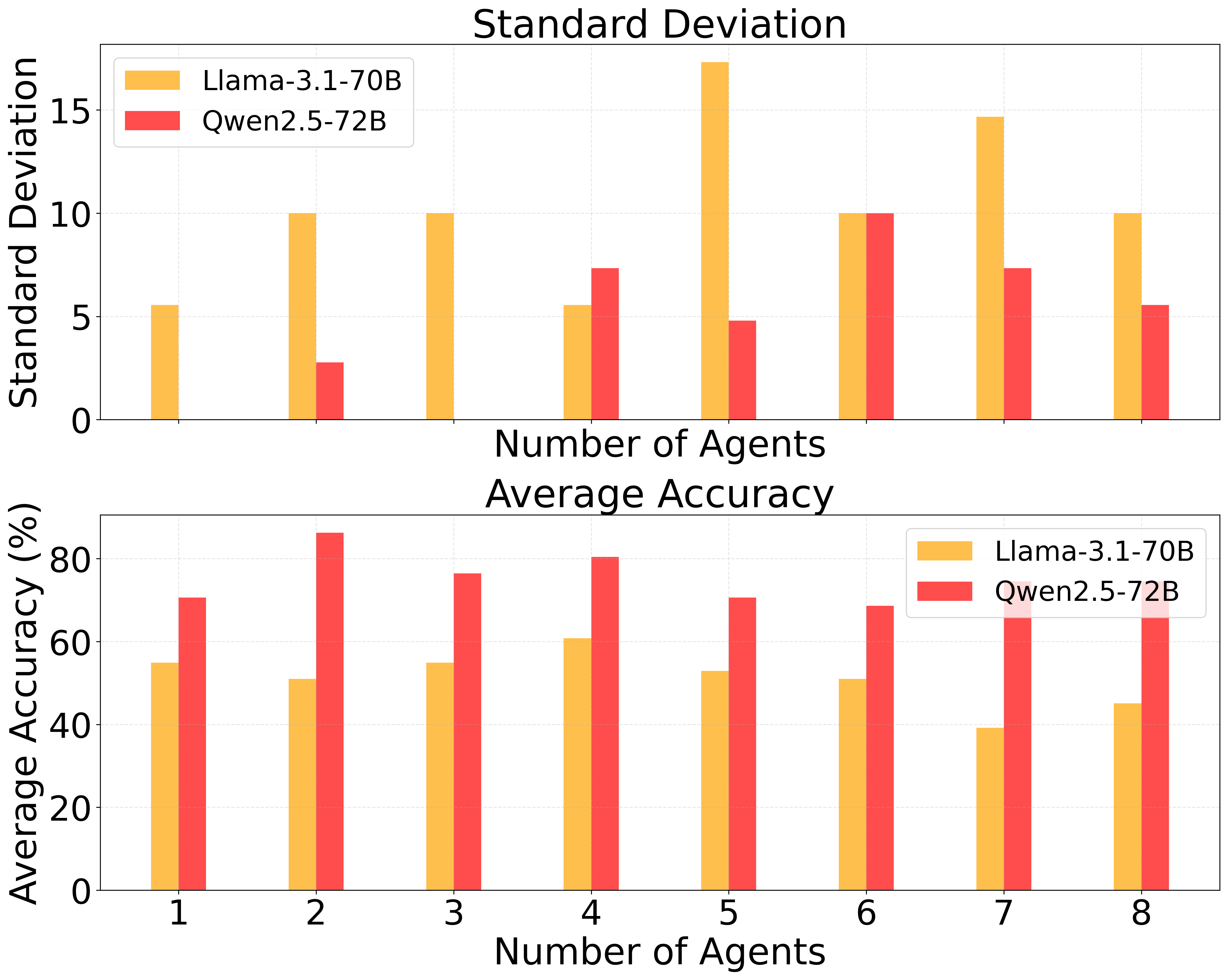}
        \caption{Model Type Impact on College Physics for Models with Large Parameters}
        \label{fig:model_type1}
\end{figure}

\begin{finding}{2}
\textbf{Task type dictates the optimal agent number size and tolerance for scaling.}
\end{finding}

We use Llama-3.1-70B and Qwen-2.5-72B for analyzing task-type effects.
Figures~\ref{fig:task_type1} to~\ref{fig:task_type3} show that scaling is fundamentally mediated by task characteristics.
In logical reasoning domains (abstract algebra, formal logic), accuracy drops substantially with more agents (27.45\% from $n{=}1$ to $n{=}8$ for Llama-3.1-70B), as reasoning chains are sensitive to information redundancy and inconsistent reasoning paths.
Conversely, fact retrieval tasks (global facts, philosophy) exhibit more complex patterns where initial declines can be partially mitigated at higher agent counts through knowledge diversity.
Notably, Llama-3.1-70B on Philosophy exhibits a U-shaped recovery at larger $n$—a phenomenon driven by redundant majority consensus statistically suppressing earlier noise rather than genuine collaborative intelligence, as we will elaborate with a case in Appendix~\ref{app:ushape}.
Qualitative analysis and complete results are in Appendix~\ref{app:dialogue_case_reasoning} and~\ref{app:full_tasks}.

\begin{finding}{3}
\textbf{When the number of agents in MAS is overly large, it will exhibit "pseudo stability" due to its poor performance.}
\end{finding}
Output stability generally decreases as agent count increases.
However, beyond a certain point, stability may paradoxically improve (Figures~\ref{fig:task_type1},~\ref{fig:task_type2},~\ref{fig:task_type3})—a state we term \textit{pseudo-stability}, occurring not from robust collaboration but from consistent failure.

Cross-model comparison reveals that while the inverted-U pattern is consistent across model families, the specific $n^*$, peak performance, and output stability vary (Appendix~\ref{app:model_scale}).
Ablation studies confirm that agent profile attributes shift absolute performance and stability but do not alter the fundamental inverted-U law, underscoring that the law is governed by interaction dynamics inherent to collaboration (Appendix~\ref{app:ablation}).

\subsection{Decoupling Agent Count from Context Length}
\label{sec:context_control}

Increasing $n$ in SIMAS also lengthens the total context, 
and to decouple these factors, we pad conversation history with neutral filler tokens to a fixed total length.

\begin{finding}{4}
\textbf{Performance degradation is causally driven by coordination overhead rather than long-context failure.}
\end{finding}

\begin{figure}[t]
    \centering
    \includegraphics[width=\columnwidth]{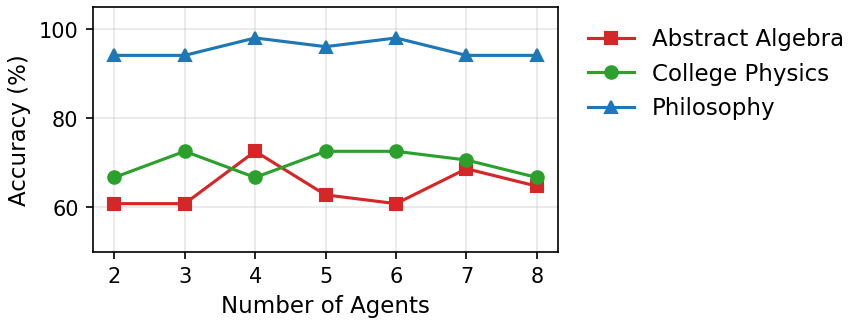}
    \caption{Qwen2.5-72B under fixed context length. The inverted-U pattern persists under fixed context length.}
    \label{fig:token_padding}
\end{figure}

As shown in Figure~\ref{fig:token_padding}, the inverted-U pattern remains clearly visible for Qwen2.5-72B across three MMLU subjects.
Abstract Algebra rises from 60.78\% ($n{=}2$) to a peak of 72.55\% ($n{=}4$) before declining to 64.71\% ($n{=}8$); Philosophy peaks at 98.04\% ($n{=}4,6$); College Physics sustains 72.55\% over $n{=}3$--6 before dropping at $n{=}8$.

\subsection{Cross-Architecture Validation with AutoGen}
\label{sec:autogen}

To test whether the inverted-U law generalizes beyond SIMAS's sequential chain topology, we implement a parallel GroupChat debate system using AutoGen~\cite{autogen}, which employs a fully connected communication topology with dynamic speaker selection.
We test Llama3.1-70B under identical conditions.

\begin{finding}{5}
\textbf{Structured debate architectures can raise peak performance and partially delay decline, but do not eliminate the fundamental inverted-U trade-off.}
\end{finding}

\begin{figure}[t]
    \centering
    \includegraphics[width=\columnwidth]{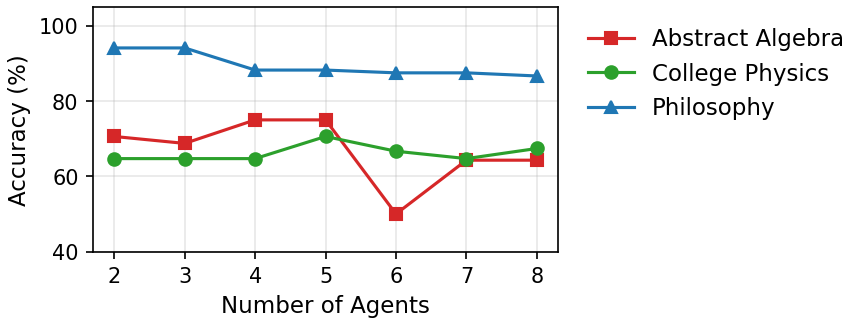}
    \caption{Llama3.1-70B under AutoGen debate. The inverted-U pattern persists under AutoGen's structured debate topology.}
    \label{fig:autogen}
\end{figure}

As shown in Figure~\ref{fig:autogen}, AutoGen achieves peak accuracies of 75.00\% on Abstract Algebra ($n{=}4,5$) and 70.59\% on College Physics ($n{=}5$), modestly higher than SIMAS's corresponding peaks.
However, the inverted-U pattern is preserved: Abstract Algebra collapses to 50.00\% at $n{=}6$, and Philosophy declines monotonically from 94.12\% ($n{=}2$) to 86.67\% ($n{=}8$).
AutoGen runs exhibited substantially lower variance than SIMAS, yet the same macroscopic pattern emerges.

\section{Comparison with Reasoning}
We now benchmark our SIMAS framework against a strong single-agent baseline: CoT prompting, which utilizes the same underlying LLM to produce a step-by-step reasoning trace followed by a final answer within a single pass.
This comparison tests whether minimalist multi-agent collaboration reliably outperforms focused single-agent reasoning from the same LLM.
Concrete prompts are provided in Appendix~\ref{app:cot}.

\begin{table*}[t]
\centering
\begin{tabular}{lccccccc}
\hline
\textbf{Architecture} & \textbf{Model} & \textbf{GSM8K} & \textbf{MATH (L1--L5)} & \textbf{AIME 2024} \\
\hline
Naive-CoT & Qwen2.5-72B & 75.13 & 77.65--63.68 & 10.0 \\
& Llama-3.1-70B & 87.33 & 88.82--80.77 & 16.7 \\
\hline
ReAct & Qwen2.5-72B & 52.76 & 35.88--28.21 & 3.3 \\
& Llama-3.1-70B & 20.09 & 11.76--4.27 & 0.0 \\
\hline
AutoGen & Qwen2.5-72B & 81.80 & 85.29--71.79 & 16.7 \\
& Llama-3.1-70B & 85.21 & 92.94--82.48 & 20.0 \\
\hline
Multi-Agent Debate & Qwen2.5-72B & 75.81 & 79.41--66.24 & 16.7 \\
& Llama-3.1-70B & 90.82 & 92.94--85.90 & 20.0 \\
\hline
\end{tabular}
\caption{Performance comparison of different reasoning and collaboration architectures across mathematical benchmarks. The data is from ~\cite{gu2025agentgroupchatv2}. Results demonstrate that sophisticated multi-agent architectures (AutoGen, Debate) can surpass single-agent CoT, while simpler paradigms (ReAct, SIMAS) struggle.}
\label{tab:arch_comparison}
\end{table*}

\begin{finding}{6}
\textbf{For reasoning tasks, the primary failure mode of minimalist MAS is the fragmentation of coherent thought.}
\end{finding}

\begin{figure}[t]
        \centering
        \includegraphics[width=\columnwidth]{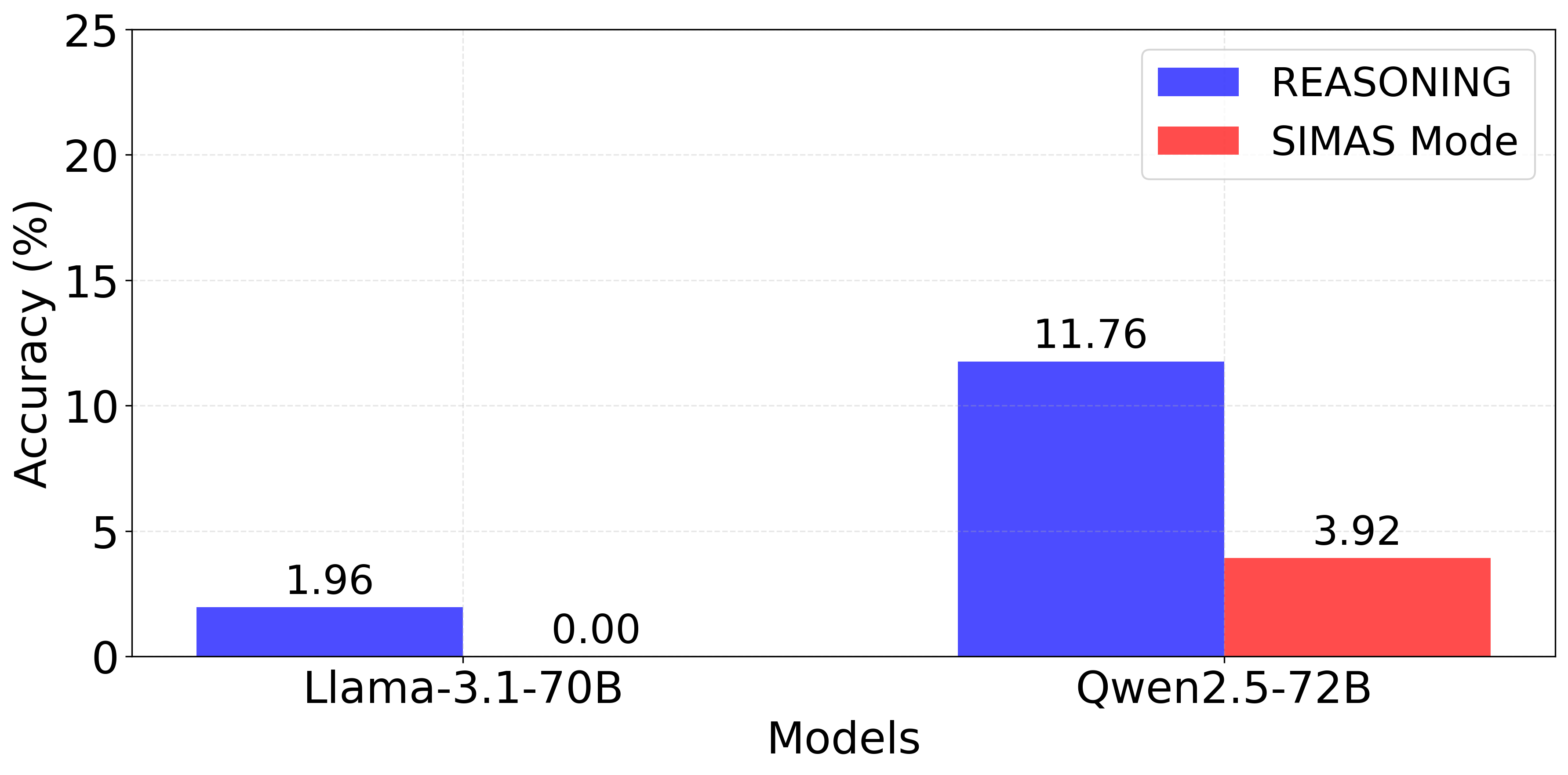}
        \caption{CoT vs. SIMAS on AIME 2025. SIMAS fails catastrophically on complex multi-step reasoning.}
        \label{fig:cot_vs_simas_aime}
\end{figure}

On reasoning-intensive benchmarks such as abstract algebra (Figure.~\ref{fig:cot_vs_simas_algebra_app}) and AIME 2025 (Figure.~\ref{fig:cot_vs_simas_aime}), CoT consistently matches or significantly surpasses the best SIMAS configuration.
The performance gap widens with problem complexity, with SIMAS failing catastrophically on AIME problems.
This failure mode stems from the fragmentation of coherent reasoning chains across sequential agent turns: without a dedicated mechanism to synthesize fragments into a unified line of reasoning or to correct accumulating errors, the collective output becomes disjointed and drifts from the correct solution path.

\begin{figure}[t]
    \centering
    \includegraphics[width=\columnwidth]{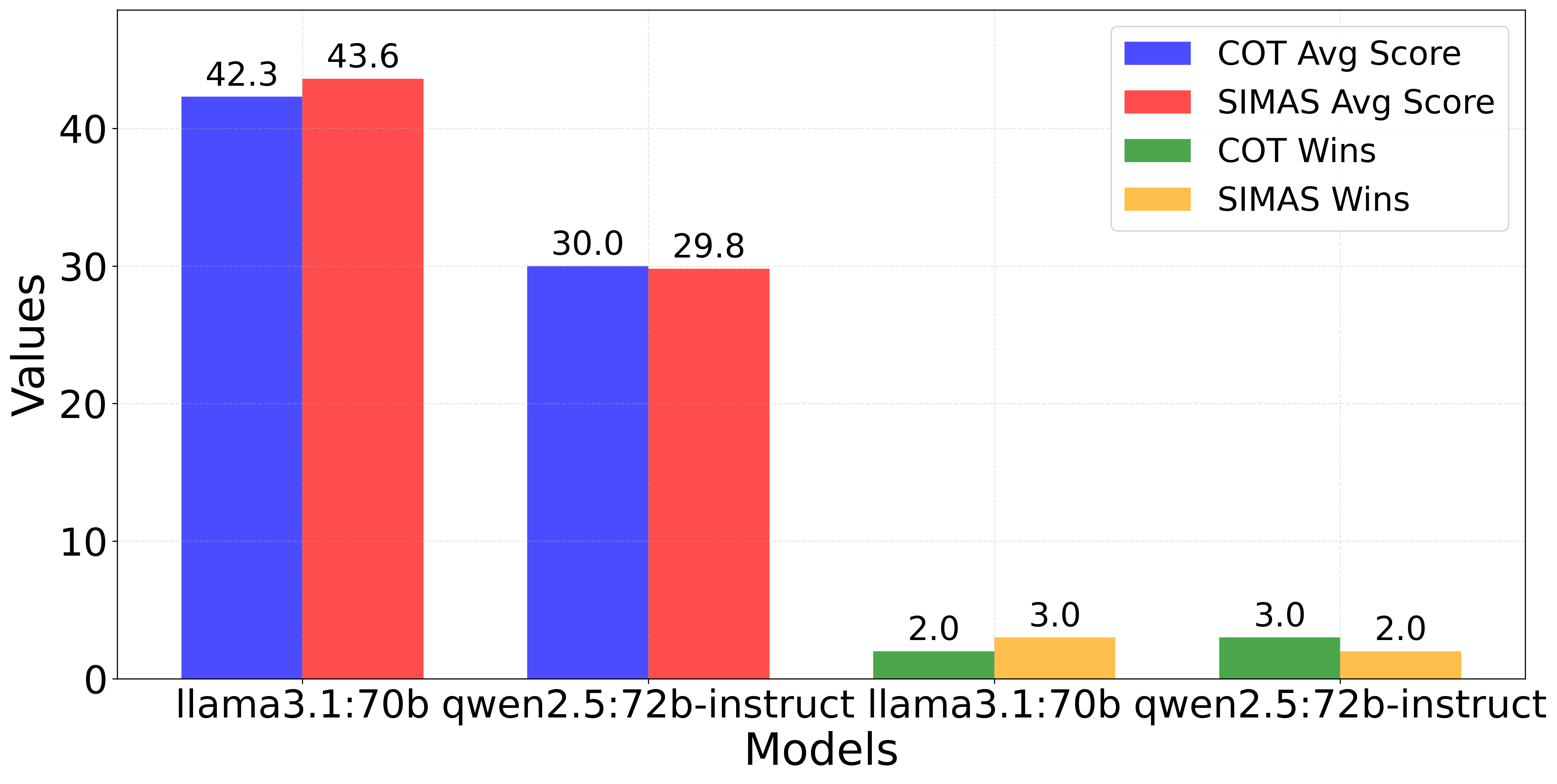}
    \caption{Win-rate comparison on coding tasks. SIMAS shows a model-dependent, marginal advantage.}
    \label{fig:code_comparison}
\end{figure}

On open-ended generation and coding tasks (Figures~\ref{fig:open_ended_comparison_app},~\ref{fig:code_comparison}), results are mixed and model-dependent: SIMAS achieves marginal advantages in some settings but shows inconsistent benefits.
Detailed evaluation protocols, qualitative case studies, and complete results are in Appendix~\ref{app:gen_eval},~\ref{app:dialogue_case_reasoning},~\ref{app:dialogue_case}.

\begin{finding}{7}
\textbf{The value of multi-agent collaboration is not automatic but is engineered through specific interaction architectures.}
\end{finding}
The frequent underperformance of SIMAS versus CoT clarifies that collective intelligence is an emergent property of \textit{interaction architecture}, not a guaranteed outcome of agent plurality.
As summarized in Table~\ref{tab:arch_comparison}, frameworks like AutoGen~\cite{autogen} and Multi-Agent Debate~\cite{liang2023debate} incorporating structured workflows and critical evaluation loops surpass single-agent CoT on challenging benchmarks (e.g., AIME 2024: 16.7--20.0\% vs.\ 16.7\% Naive-CoT with Llama-3.1-70B), while SIMAS, lacking such mechanisms, incurs overhead without reliably capturing synergistic benefits.
This, together with our AutoGen replication (Section~\ref{sec:autogen}), implies that the pursuit of collective intelligence must be architectural, not merely numerical.

\section{Conclusion}
This work systematically demonstrates that LLM-based Multi-Agent System performance does not scale linearly with agent count but exhibits a pattern of diminishing returns. Effective collaboration first requires a sufficiently capable base LLM. The optimal number of agents is a critical design parameter, heavily dependent on task type and model architecture, balancing synergy against overhead. The performance degradations derived from collaboration overhead generalizes across interaction architectures. Crucially, collective intelligence is not an automatic outcome of adding agents but an emergent property contingent on deliberate interaction design. Without architectural support for synthesis and refinement, multi-agent dialogue risks inefficiency. Future MAS development must therefore prioritize designing adaptive, task-aware collaboration protocols over simply increasing agent plurality.

\section*{Limitations}
This study, while providing foundational insights into MAS scaling behavior, has several limitations that point to future research directions.
First, our focus on homogeneous agents using a single LLM isolates collaboration effects but does not capture the heterogeneous model and tool use common in practical systems, where complementarity might alter scaling dynamics.
Second, our minimalist SIMAS architecture, while clarifying first principles, omits sophisticated coordination mechanisms (e.g., voting, dynamic workflows) that could mitigate the overhead we observed.
While our AutoGen comparison (Section~\ref{sec:autogen}) confirms the inverted-U law's cross-architectural generality, exploring a broader range of topologies (DAG-based, hierarchical) and their interaction with model capability remains an open direction.
Third, although our token-padding experiment (Section~\ref{sec:context_control}) confirms that coordination overhead rather than long-context failure drives the inverted-U pattern, we evaluated only output accuracy; deeper psycholinguistic analysis of agent dialogues---quantifying semantic diversity, conversational dynamics, and belief divergence---could further characterize the micro-level mechanisms of collaboration breakdown.
Fourth, our evaluation primarily uses closed-book QA benchmarks, which may not fully capture collaborative benefits in longitudinal, creative, or tool-augmented tasks.
Finally, scaling was tested only up to 8 agents; the dynamics of much larger collectives and their potential for novel emergent phenomena remain unexplored.

\section*{Ethical Concerns}
This study focuses on the algorithmic and architectural principles of multi-agent scaling, utilizing exclusively publicly available and widely-adopted benchmark datasets (e.g., MMLU, AIME, E-KAR). 
No sensitive personal data, simulated social dynamics, or real-world decision-making scenarios are involved. The Sequential Iterative Multi-Agent System (SIMAS) framework is designed as a minimalist research tool to investigate fundamental collaboration dynamics, explicitly excluding applications in opinion manipulation, autonomous action with real-world consequences, or the generation of deceptive content. 
All experimental interactions are confined to closed, controlled environments for problem-solving, and no user privacy information is stored in our codebase or logs; only aggregated performance metrics are reported.
During the research process, AI assistants (including DeepSeek-R1) were employed as tools to aid in specific auxiliary tasks such as brainstorming initial ideas, generating code for experimental pipelines, polishing textual descriptions, and creating visualizations. We explicitly state that all core research ideas, experimental design, data analysis, interpretation of results, and scientific conclusions were originated, critically evaluated, and decisively finalized by the human authors. The AI tools served strictly as supportive instruments, and every piece of their output was rigorously reviewed, validated, and often substantially revised by the authors to ensure correctness and alignment with the research objectives.
The primary ethical considerations—such as potential reasoning bias, coherence fragmentation, or the amplification of base model errors—are emergent properties stemming from the underlying LLM capabilities and the chosen task prompts, rather than inherent flaws of the SIMAS architecture itself. 
Should the principles explored here inform the design of future MAS deployed in open or user-facing contexts, rigorous additional risk assessment and governance mechanisms would be mandatory, adhering strictly to established ethical guidelines in AI research and development.

\bibliography{references}

\appendix
\section{Formal Definitions for MAS}
\label{app:formal_defs}

\begin{definition}[System Output]
For an input question $q \in \mathcal{Q}$, the system output is defined as:
\[
o = f_{\mathcal{M}}(q) = \Pi_{output}(h_{n,T}, a_1)
\]
where $h_{n,T} \in \mathcal{H}$ represents the complete conversation history after $T$ rounds of discussion.
\end{definition}

\begin{definition}[Agent Configuration]
Each agent $a_i \in \mathcal{A}$ is configured as a quadruple:
\[
a_i = \langle \text{LLM}, d_i, p_i, b_i, e_i \rangle
\]
where $\text{LLM}$ represents the underlying language model (shared across all agents),
$d_i \in \mathcal{D}$ denotes the description,
$p_i \in \mathcal{P}$ denotes the personality trait,
$b_i \in \mathcal{B}$ denotes the core beliefs,
$e_i \in \mathcal{E}$ denotes the expertise,
with $\mathcal{D}, \mathcal{P}, \mathcal{B}, \mathcal{E}$ representing the respective attribute spaces.
\end{definition}

\begin{property}[Model Homogeneity]
All agents share the same underlying language model:
\[
\forall a_i, a_j \in \mathcal{A}, \quad \text{LLM}(a_i) = \text{LLM}(a_j)
\]
\end{property}

\begin{property}[Configuration Heterogeneity]
Agents achieve diversity through distinct configuration parameters:
\[
\forall a_i, a_j \in \mathcal{A}, i \neq j \Rightarrow (d_i, p_i, b_i) \neq (d_j, p_j, b_j)
\]
\end{property}

\section{Framework of SIMAS}

\subsection{Problem Provider}
The \textbf{Problem Provider} module loads datasets (e.g., MMLU, E-KAR), preprocesses questions to ensure consistency, and normalizes answers by mapping responses to standardized formats (e.g., options A/B/C/D).

\subsection{Agent Generator}
The \textbf{Agent Generator} takes as inputs the discipline type of the problem and the number of agents \( N \), and outputs \( N \) dynamically generated agents. These agents are created by invoking the LLM API with tailored prompts to assign unique roles and personalities.

\subsection{Group Chat Engine}
The \textbf{Group Chat Engine} executes the main workflow of SIMAS, managing multi-round discussions. It monitors for timeouts or errors and implements a retry mechanism with a maximum of 100 retries (\texttt{max\_retries=100}).

\subsection{Evaluator}
The \textbf{Evaluator} compares the final group chat answer with the ground truth and records auxiliary metrics, such as confidence scores, to assess response quality.

The framework leverages the LLM's API for text generation and processing, offering the following key functionalities: It supports \textbf{Agent Generation} by automatically creating agents with distinct personalities and expertise based on specified quantities and role requirements. Its \textbf{Group Chat Management} orchestrates communication among agents, ensuring sequential contributions and maintaining discussion history for context-aware responses. The \textbf{Answer Evaluation} functionality extracts the final answer from the last agent's response after group chat completion and compares it with the ground truth to compute accuracy.

The framework supports seamless switching between different language models, enabling comparative experiments across models. This flexibility ensures robust evaluation of multi-agent system performance under varying configurations.

\section{Mechanisms of U-Shaped Recovery}
\label{app:ushape}

A notable deviation from the canonical inverted-U pattern occurs in Philosophy with Llama3.1-70B, where accuracy exhibits a U-shaped curve: declining from the single-agent baseline through intermediate agent counts before partially recovering at $n{=}8$. Qualitative analysis of agent dialogues on an MMLU item concerning Aristotle's concept of \textit{incontinence} (knowing the wrong and doing it anyway) reveals a three-phase mechanism.

In Phase~1 ($n{=}2$--3), dialogue remains focused and efficient: one agent dominates with correct reasoning while coordination overhead is minimal. In Phase~2 ($n{=}4$--5), social meta-discourse (apologies, acknowledgments, redundant praise) proliferates and divergent minority interpretations emerge---for instance, conflating \textit{akrasia} with the distinct concept of \textit{enkrateia}. This fragments the reasoning chain and causes the accuracy dip. In Phase~3 ($n{=}8$), the sheer volume of majority agents reiterating the correct pre-training-consistent answer statistically suppresses earlier noise, producing an apparent recovery. This recovery, however, constitutes \textit{pseudo-stability} rather than genuine collaborative intelligence, as it depends on the model's pre-training bias toward the consensual answer rather than on structured synthesis of diverse perspectives. Qwen2.5-72B does not exhibit the U-shape, attributable to its lower semantic diversity and stronger long-context robustness.

This case reveals a key distinction: \textbf{error-propagating tasks} (mathematics, formal logic), where a single incorrect step irreversibly derails the solution, exhibit classic inverted-U patterns with steep post-peak declines; \textbf{consensus-recoverable tasks} (interpretive philosophy, fact retrieval with high prior agreement), where majority voting over redundant correct responses can mask mid-scale coordination failures, may exhibit U-shaped recovery contingent on model-specific semantic diversity.

\section{Supplementary Material for Experiments}
\label{app:exp}

\subsection{Agent Profile Generation and Examples}
\label{app:agent_profiles}
Profiles were generated by prompting the base LLM with a template to create diverse sets of $\langle personality, core\ belief, expertise \rangle$ triples. For a 4-agent team on a reasoning task, examples include:
\begin{verbatim}
Agent 1: ("Skeptical", 
"Precision outweighs speed", 
"Logical Verification")

Agent 2: ("Creative", 
"Novel approaches are valuable", 
"Alternative Solution Generation")
...
\end{verbatim}

The prompt template used for agent profile generation is presented below.
\vspace{\baselineskip} 

\begin{ACLPrompt} 
\textbf{System Instruction:}\\
Generate a description in English for an AI agent that will participate in a group discussion to solve problems of \textit{\{subject\}}. This is agent \textit{\{index\}} of \textit{\{total\}} total agents.

\textbf{Context Requirements:}\\
Existing agent names: \textit{\{name\}}\\
Existing agent descriptions: \textit{\{description\}}

\textbf{Output Structure (JSON format):}\\
Please provide the following:\\
1. \textbf{name}: A simple name for the agent (unique from existing agents)\\
2. \textbf{personality}: One of [\texttt{FRIENDLY, SKEPTICAL, NEUTRAL, AGGRESSIVE}]\\
3. \textbf{description}: The i-th participant in the chat.\\
4. \textbf{expertise}: The agent's main strength in problem solving of \textit{\{subject\}}\\
5. \textbf{beliefs}: List of 3-5 beliefs relevant to problem solving of \textit{\{subject\}} that this agent holds.

\textbf{Example Output:}\\
{\ttfamily \small 
\{ \\
\hspace*{4mm}"name": "Alice", \\
\hspace*{4mm}"personality": "FRIENDLY", \\
\hspace*{4mm}"description": "A warm and friendly assistant who excels at explaining complex concepts through diagrams", \\
\hspace*{4mm}"expertise": "visualization and explanation", \\
\hspace*{4mm}"beliefs": [ \\
\hspace*{8mm}"Clear visualization aids understanding", \\
\hspace*{8mm}"Multiple approaches should be considered", \\
\hspace*{8mm}"Collaboration improves problem solving" \\
\hspace*{4mm}] \\
\}
}

\vspace{1ex}
\textbf{Important Constraints:}\\
Remember to generate descriptions in English. Make sure each agent has unique beliefs that align with their role and expertise. Make sure your output is and only is valid JSON format, without any other words before or after the JSON content. You must generate the last agent with the personality \texttt{AGGRESSIVE} and relevant description.
\end{ACLPrompt}

\captionof{figure}{Prompt template for agent profile generation.}
\label{fig:agent-generation}

\vspace{\baselineskip} 

\subsection{Model-Scale Scaling Results on College Physics}
\label{app:model_scale}
\begin{figure}[h]
    \centering
    \includegraphics[width=\columnwidth]{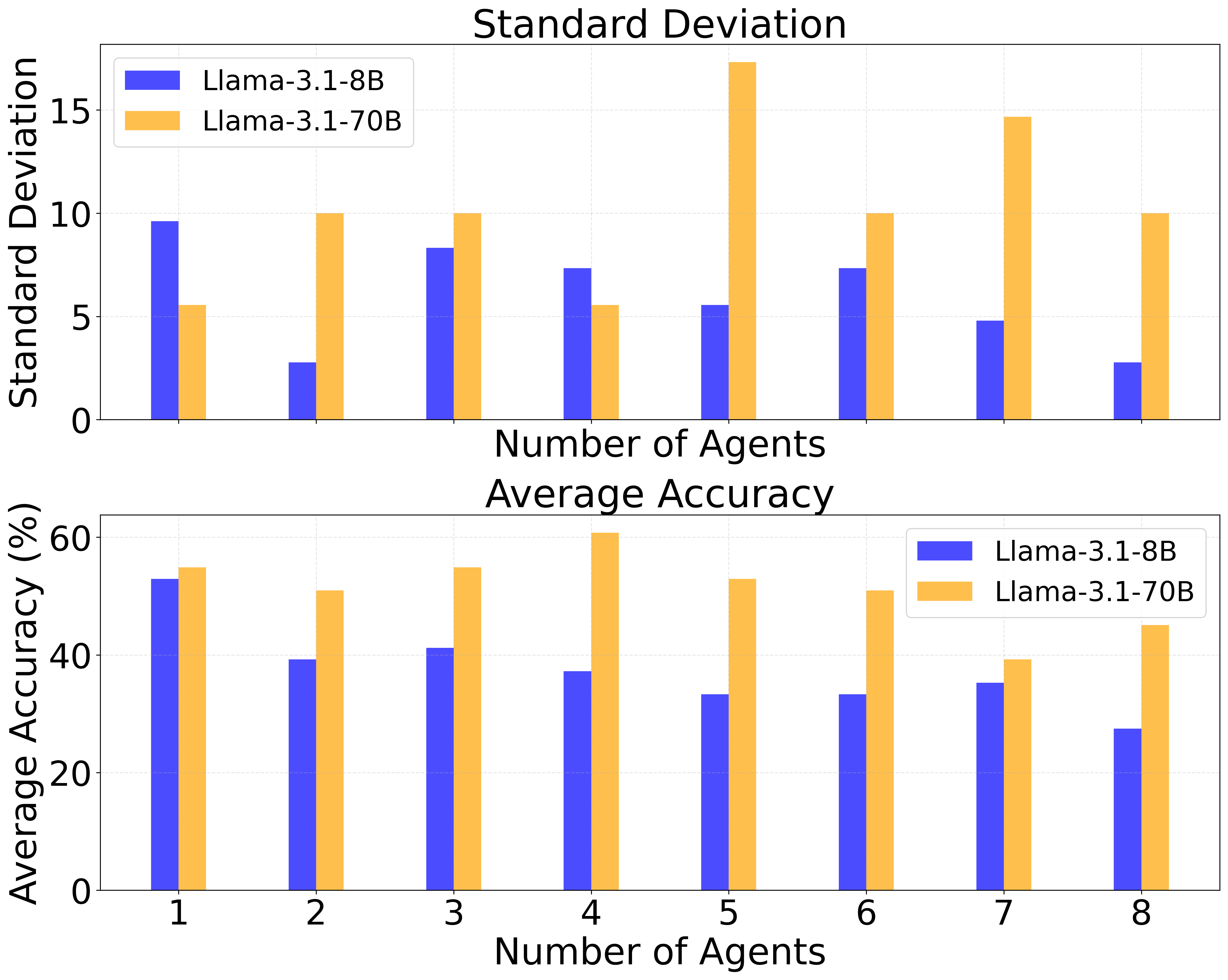}
    \caption{Model Scale Impact of Llama 3.1 on College Physics}
    \label{fig:model_scale3}
\end{figure}

\begin{figure}[h]
    \centering
    \includegraphics[width=0.9\columnwidth]{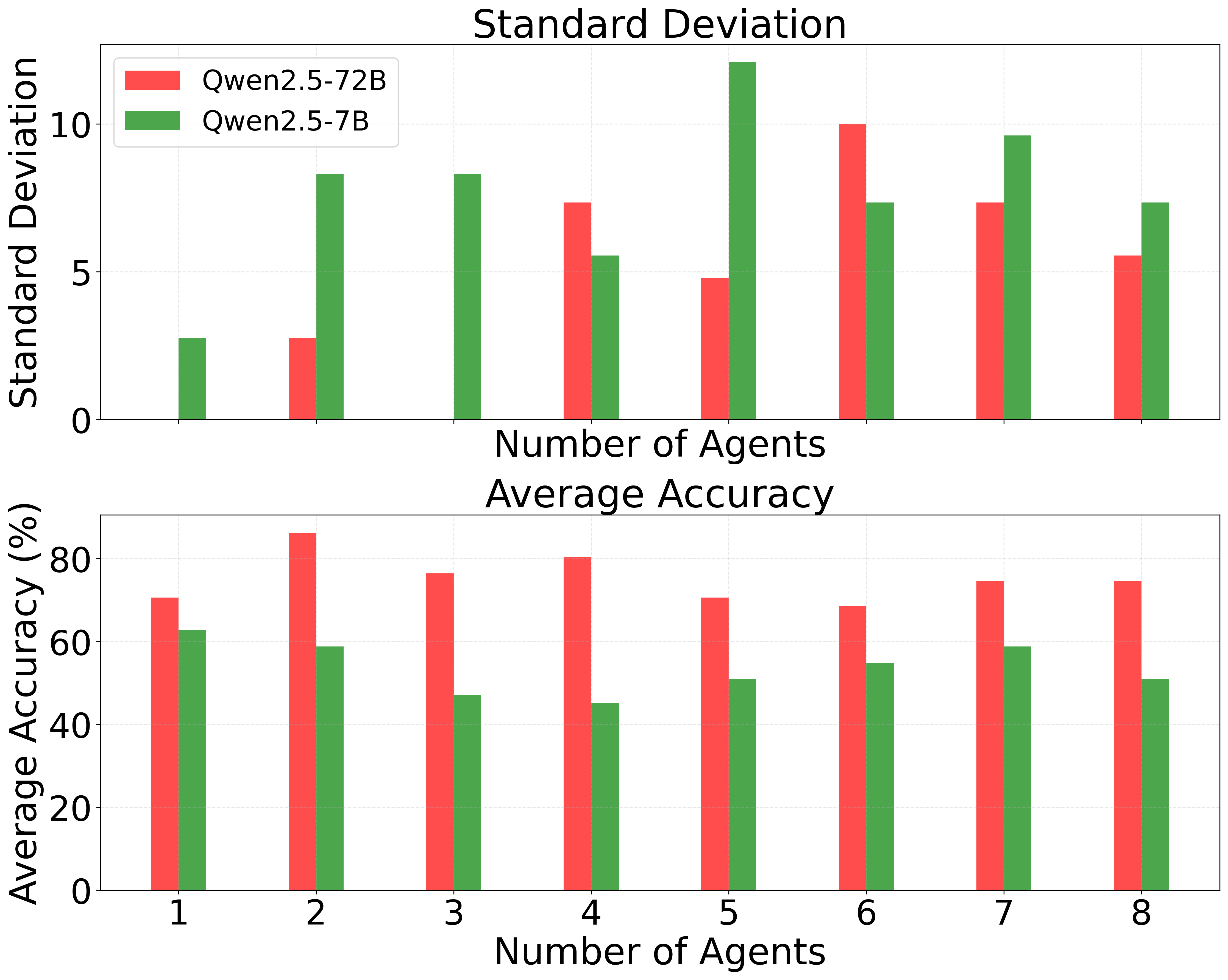}
    \caption{Model Scale Impact of Qwen2.5 on College Physics}
    \label{fig:model_scale4}
\end{figure}

\begin{finding}{1}
\textbf{Model scale provides necessary conditions for effective multi-agent collaboration, but does not ensure stable performance gains alone.}
\end{finding}
Crucially, comparing Figure~\ref{fig:model_scale1} and Figure~\ref{fig:model_scale2}, it is obviously revealed that smaller models (7B/8B parameters) proved fundamentally inadequate for multi-agent collaboration, showing monotonic performance degradation with additional agents, suggesting a minimum capability threshold for effective multi-agent interactions. 
While larger models consistently outperformed their smaller counterparts in absolute accuracy, this advantage did not uniformly extend to stability metrics. 
The Qwen2.5-72B model demonstrated also high stability, but in contrast, Llama-3.1-70B showed significant volatility despite its parameter advantage, with standard deviations reaching 17.32\%, indicating that scale alone cannot ensure consistent multi-agent performance. (see Figure~\ref{fig:model_scale3} and ~\ref{fig:model_scale4} in Appendix~\ref{app:model_scale})

\subsection{Full Task-Type Scaling Results}
\label{app:full_tasks}
\begin{table}[h]
\centering

{\footnotesize  
\begin{tabular}{p{0.25\textwidth}p{0.18\textwidth}}
\toprule
\textbf{Dataset/Discipline} & \textbf{Task Type} \\
\midrule
Abstract Algebra & Logical Reasoning \\
Formal Logic & Logical Reasoning \\
Philosophy & Fact Retrieval \\
Global Facts & Fact Retrieval \\
College Physics & Mixed \\
College Computer Science & Mixed \\
\midrule
E-KAR & Analogical Reasoning \\
\bottomrule
\end{tabular}
}
\caption{Datasets and Task Characteristics}
\label{tab:subject_compare}
\end{table}

\begin{figure*}[t]
    \centering
    \begin{subfigure}[b]{0.3\textwidth}
        \centering
        \includegraphics[width=\textwidth]{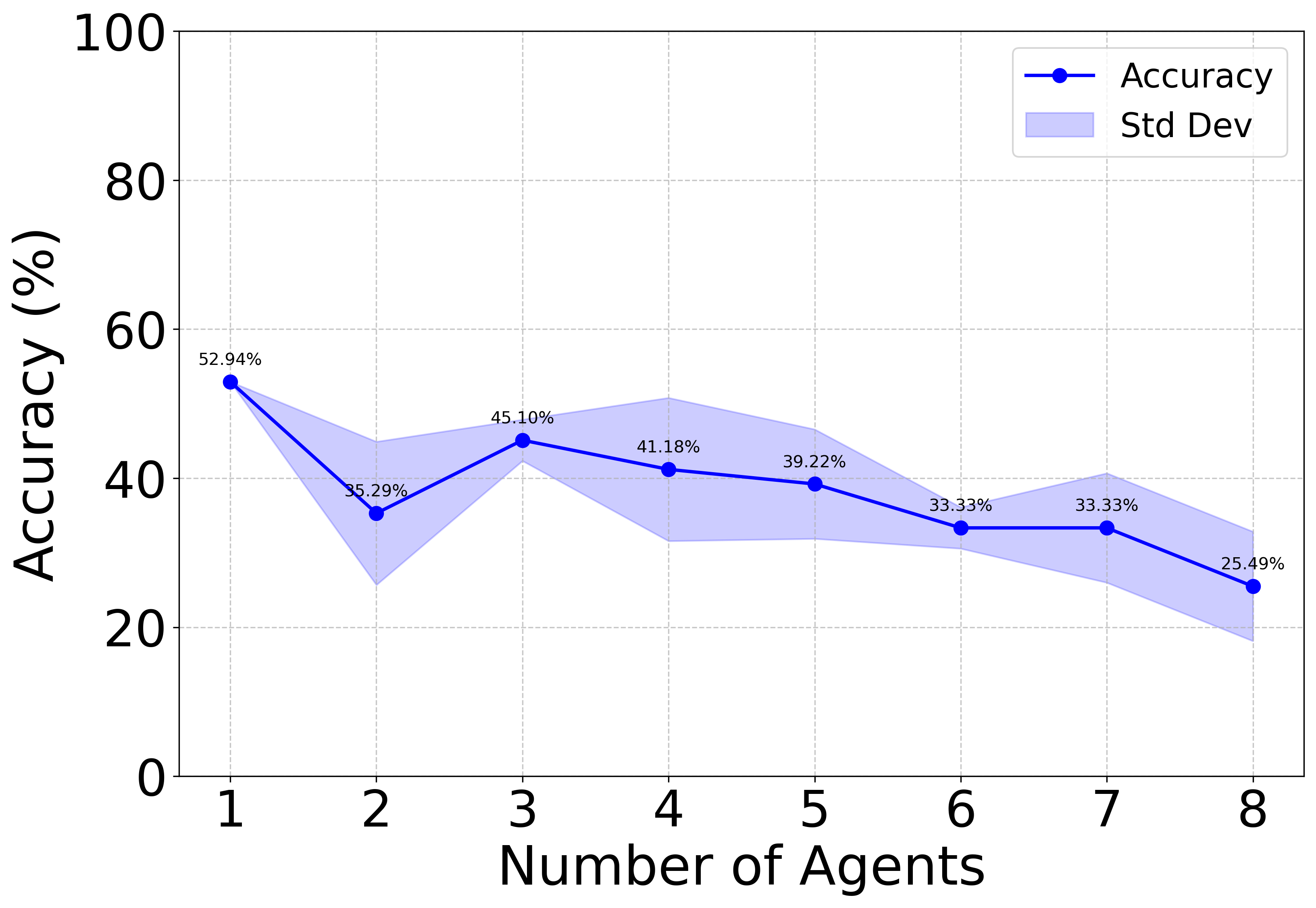}
        \caption{Performance on Abstract Algebra}
        \label{fig:task_type1_app}
    \end{subfigure}
    \hfill
    \begin{subfigure}[b]{0.3\textwidth}
        \centering
        \includegraphics[width=\textwidth]{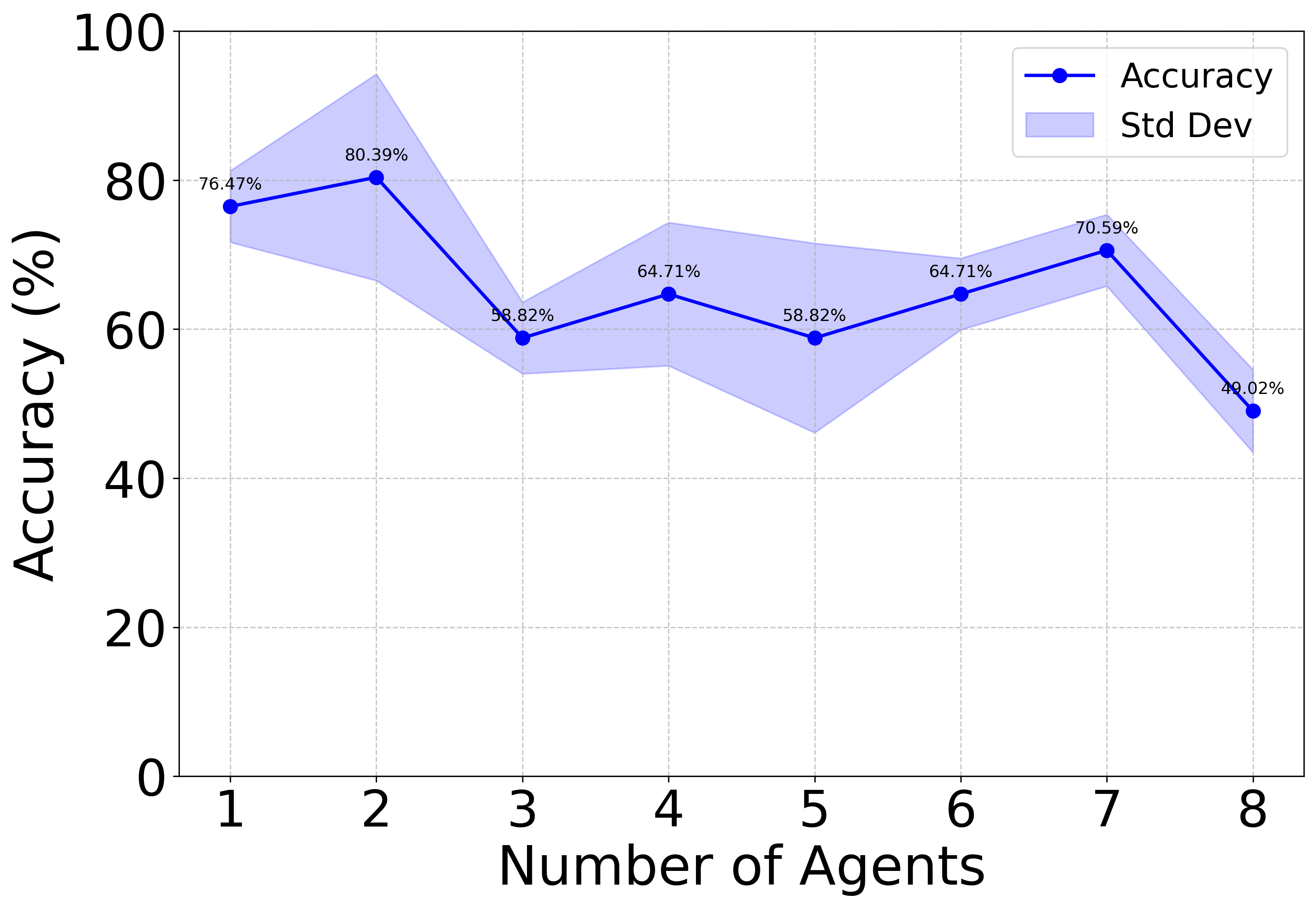}
        \caption{Performance on Philosophy}
        \label{fig:task_type2_app}
    \end{subfigure}
    \hfill
    \begin{subfigure}[b]{0.3\textwidth}
        \centering
        \includegraphics[width=\textwidth]{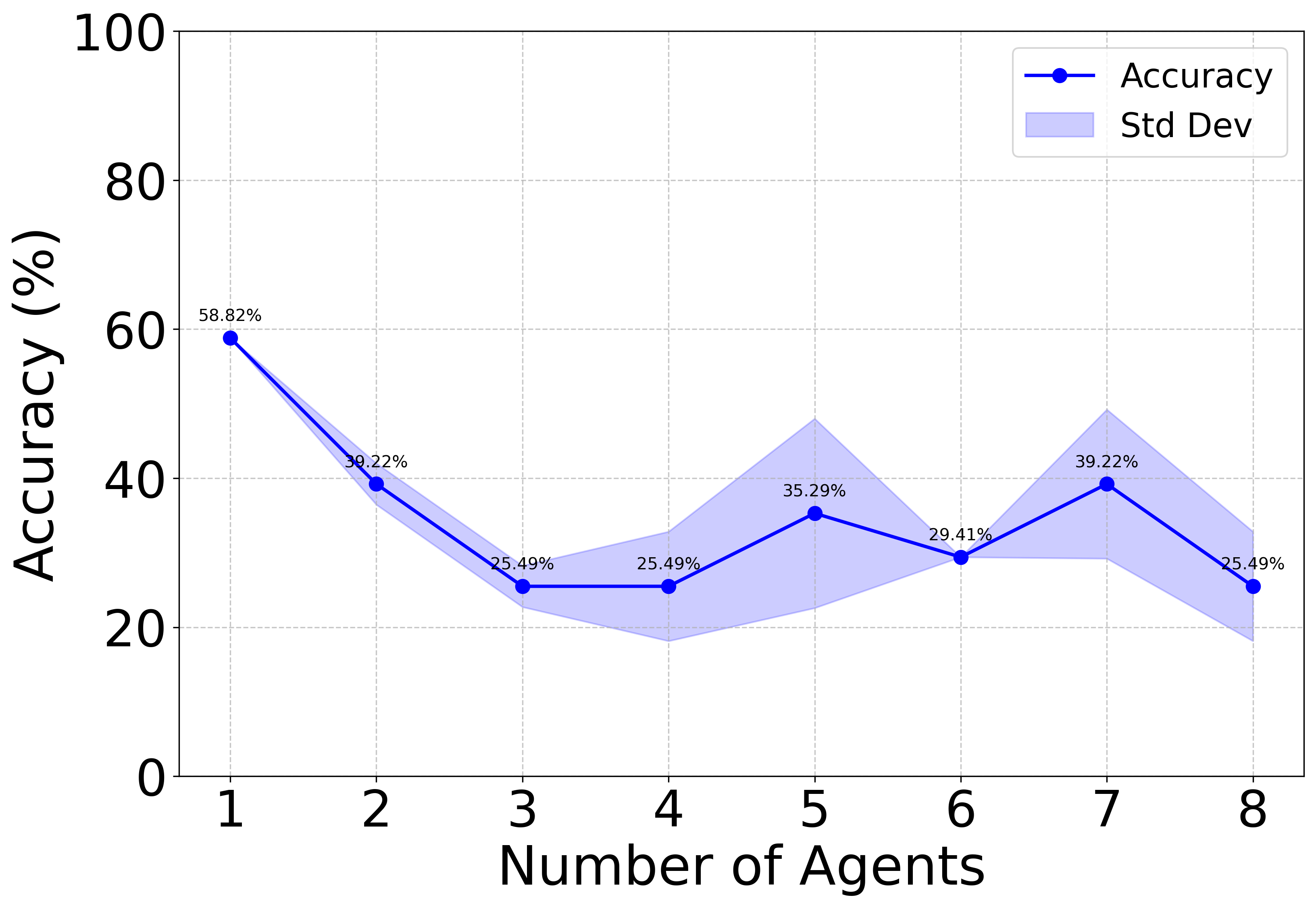}
        \caption{Performance on Formal Logic}
        \label{fig:task_type3_app}
    \end{subfigure}
    
    \vspace{0.3cm} 
    
    \begin{subfigure}[b]{0.3\textwidth}
        \centering
        \includegraphics[width=\textwidth]{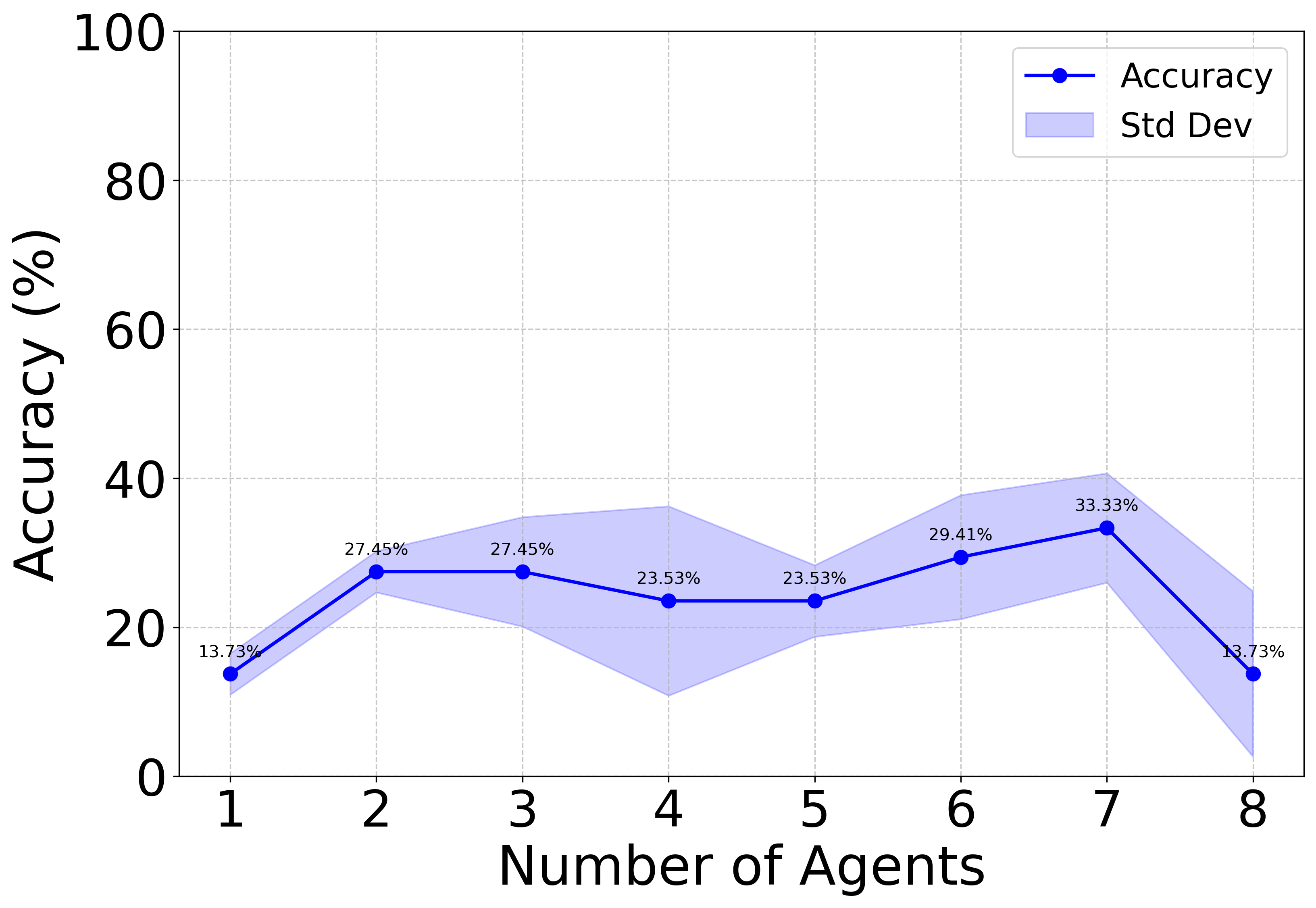}
        \caption{Performance on Global Facts}
        \label{fig:task_type4_app}
    \end{subfigure}
    \hfill
    \begin{subfigure}[b]{0.3\textwidth}
        \centering
        \includegraphics[width=\textwidth]{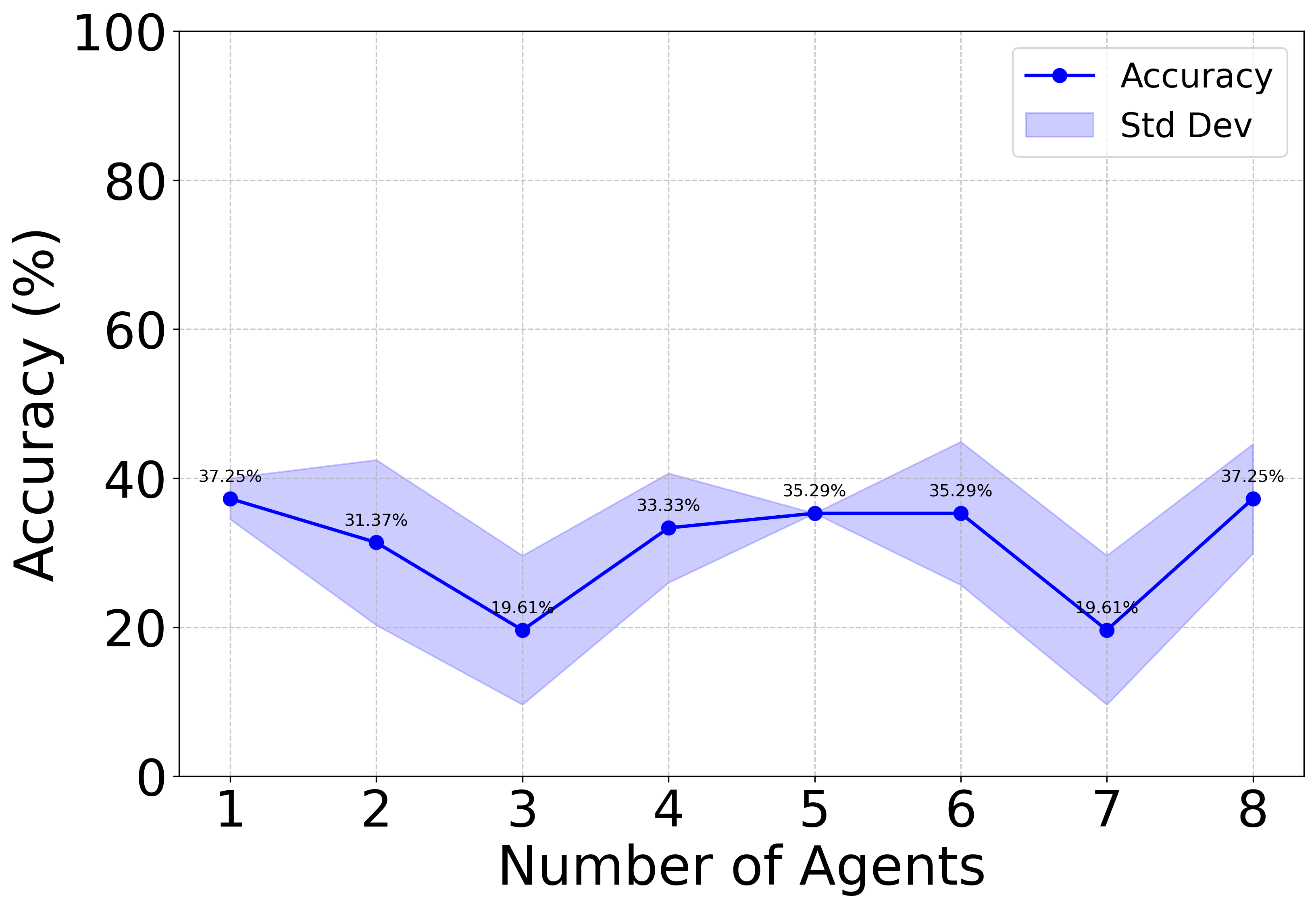}
        \caption{Performance on E-kar}
        \label{fig:task_type5_app}
    \end{subfigure}
    \hfill
    \begin{subfigure}[b]{0.34\textwidth}
        \centering
        \includegraphics[width=\textwidth]{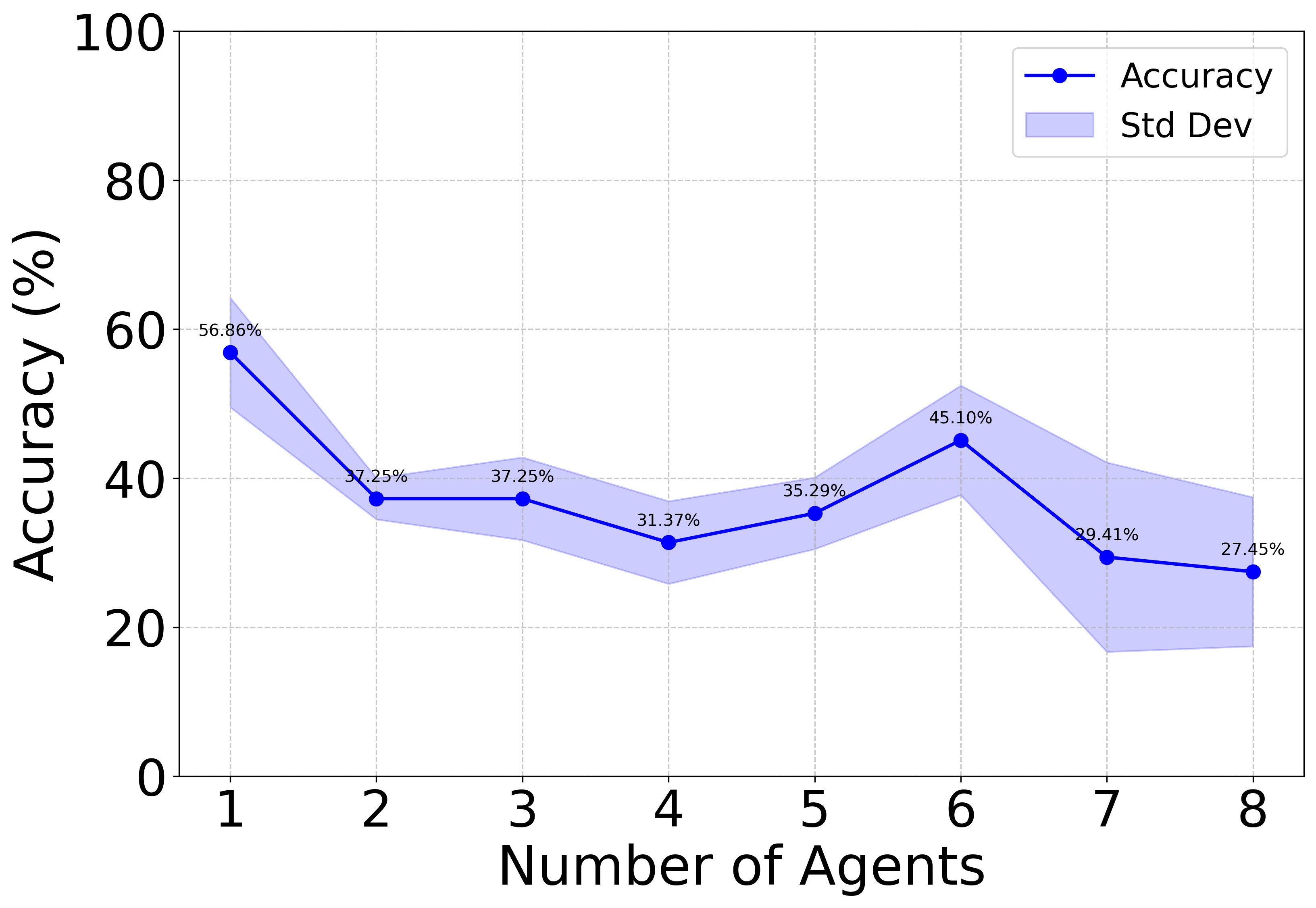}
        \caption{Performance on College Computer Science}
        \label{fig:task_type6_app}
    \end{subfigure}
    \caption{Performance across different task types on Meta-Llama-3.1-70B.}
\end{figure*}
Figures~\ref{fig:task_type1_app} to~\ref{fig:task_type6_app} illustrate that the impact of agent scaling is fundamentally mediated by task characteristics.

The experimental results reveal distinct accuracy patterns across task categories. In \textbf{logical reasoning} domains, performance trends were inconsistent. For \textbf{formal logic}, accuracy declined substantially from 58.82\% (1 agent) to 25.49\% (8 agents), a drop of 33.33 percentage points. \textbf{Abstract algebra} showed a similar overall decline from 52.94\% to 25.49\%, though with intermediate fluctuations (e.g., 45.10\% at 3 agents). This degradation stems from the inherent sensitivity of logical reasoning chains to disruptions caused by information redundancy and inconsistent reasoning paths across multiple agents.

Conversely, \textbf{fact retrieval} tasks demonstrated varied patterns. The \textbf{global\_facts} task showed low but variable accuracy (13.73\% to 33.33\%), with no clear monotonic trend relative to agent count. The \textbf{philosophy} task maintained high accuracy overall (94.12\% with 1 agent) but experienced a significant decline to 60.78\% at 4 agents before partially recovering to 74.51\% with 8 agents, suggesting that initial performance declines could be partially mitigated through knowledge diversity at higher agent counts in some domains.

\textbf{Mixed-reasoning} tasks exhibited complex, non-monotonic relationships with agent count. \textbf{College physics} accuracy fluctuated between 25.49\% and 50.98\%, peaking at 7 agents. \textbf{College computer science} showed a decline from 56.86\% (1 agent) to 27.45\% (8 agents), but with a local peak of 45.10\% at 6 agents. For analogical-reasonng, the \textbf{e-KAR} task displayed high volatility, with accuracy dropping to 19.61\% at both 3 and 7 agents.

The analysis of experimental results, incorporating standard deviation data, reveals distinct stability profiles across task types and agent counts. 

\textbf{Global Facts}, as a fact-retrieval task, shows generally low output variability for smaller agent groups, with standard deviations remaining at or below 7.34\% for one to three agents. However, stability degrades significantly at larger scales, with standard deviations reaching 12.71\% and 11.09\% for four and eight agents, respectively. This indicates that while factual recall is robust in small-group settings, coordination or consensus failures in larger multi-agent systems can introduce substantial output inconsistency.

\textbf{Formal Logic} demonstrates a pronounced instability in its reasoning chains. While performance with a single agent is perfectly stable (0.00\% std), the introduction of additional agents leads to high variability, particularly at intermediate counts. A standard deviation peak of 12.71\% at five agents confirms the inherent fragility of multi-step deductive processes in collaborative environments, where minor reasoning divergences can amplify.

\textbf{Abstract Algebra} exhibits a distinct pattern where instability is highest not at maximal agent counts but at specific intermediate configurations. Significant standard deviations of 9.61\% are observed for both two and four agents, suggesting that certain group sizes may create ambiguous task decompositions or conflict in symbolic manipulation strategies, leading to less deterministic outputs.

\textbf{College Physics}, a mixed-reasoning task, confirms a pattern of high instability at intermediate collaboration scales. Standard deviations are highest (exceeding 12\%) for two, three, and four agents, indicating that integrating conceptual knowledge with quantitative reasoning presents a critical coordination challenge. Stability does not consistently improve with more agents, as groups of six to eight maintain high variability (~10\% std).

\textbf{Philosophy}, while achieving high accuracy, shows noteworthy variability in specific multi-agent settings. The standard deviation reaches 10.00\% for four agents, suggesting that tasks involving nuanced textual interpretation and argumentation are susceptible to diverging perspectives within a medium-sized group, even if larger groups (e.g., seven or eight agents) manage to reconverge on more stable outputs.

\textbf{College Computer Science} presents a case of escalating instability with scale for this applied reasoning domain. Standard deviations show a generally increasing trend, culminating in a high of 12.71\% for seven agents. This indicates that collaborative problem-solving on complex, structured tasks may suffer from accumulating integration errors or conflicting solution approaches as the number of contributing agents grows.

\textbf{E-KAR} demonstrates that instability is not monotonic with agent count. High standard deviations of 11.09\% and 10.00\% for two and three agents, respectively, drop to 0.00\% for five agents before rising again. This non-linear pattern suggests the existence of specific, potentially task-dependent, agent group sizes that can either mitigate or exacerbate variability in decision-making or pattern recognition tasks.

\subsection{Model-Type Scaling Results on College Physics}
\begin{figure}[t]
    \centering
    \includegraphics[width=\columnwidth]{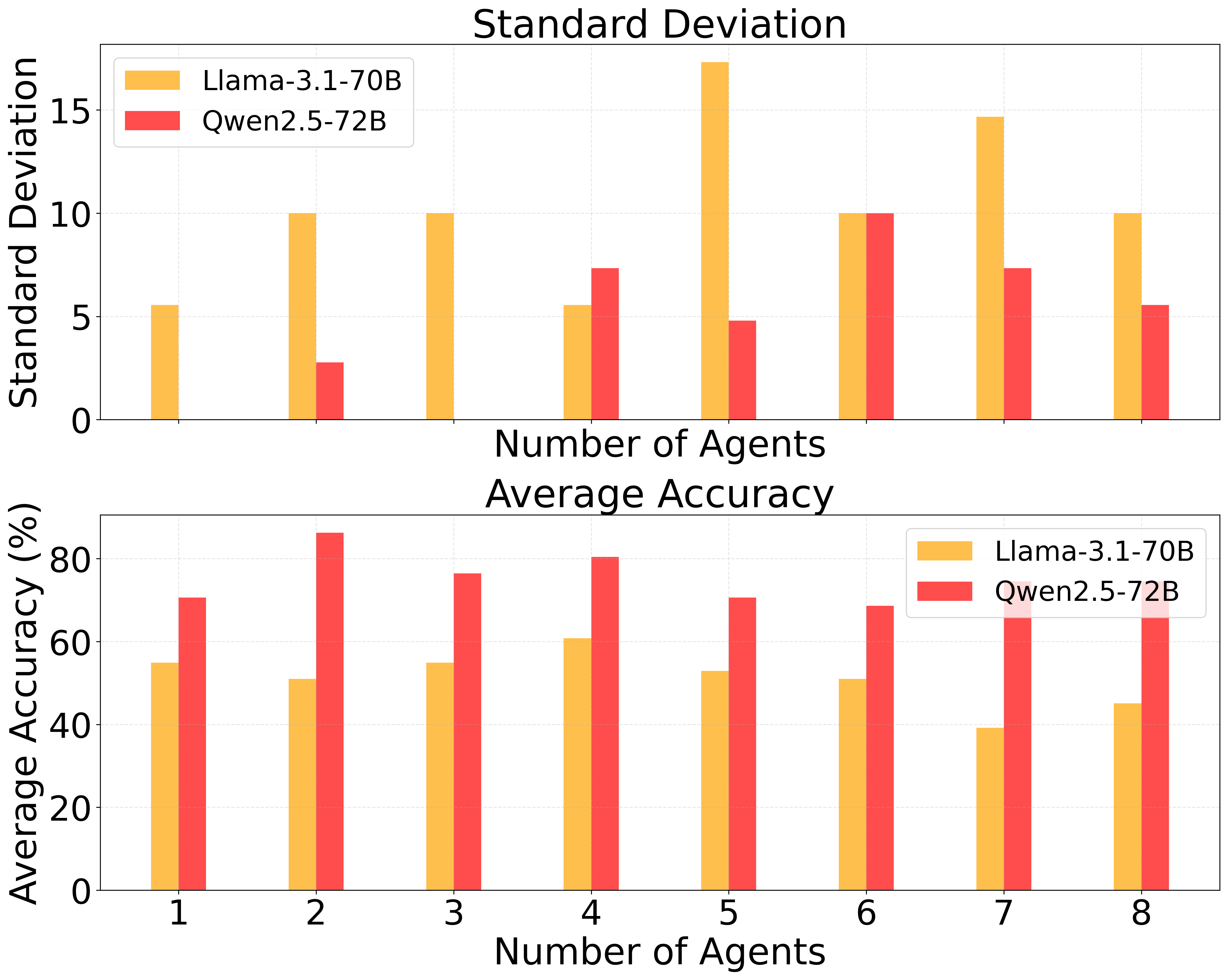}
    \caption{Model Type Impact on College Physics for Models with Large Parameters}
    \label{fig:model_type3_app}
\end{figure}

\begin{figure}[t]
    \centering
    \includegraphics[width=\columnwidth]{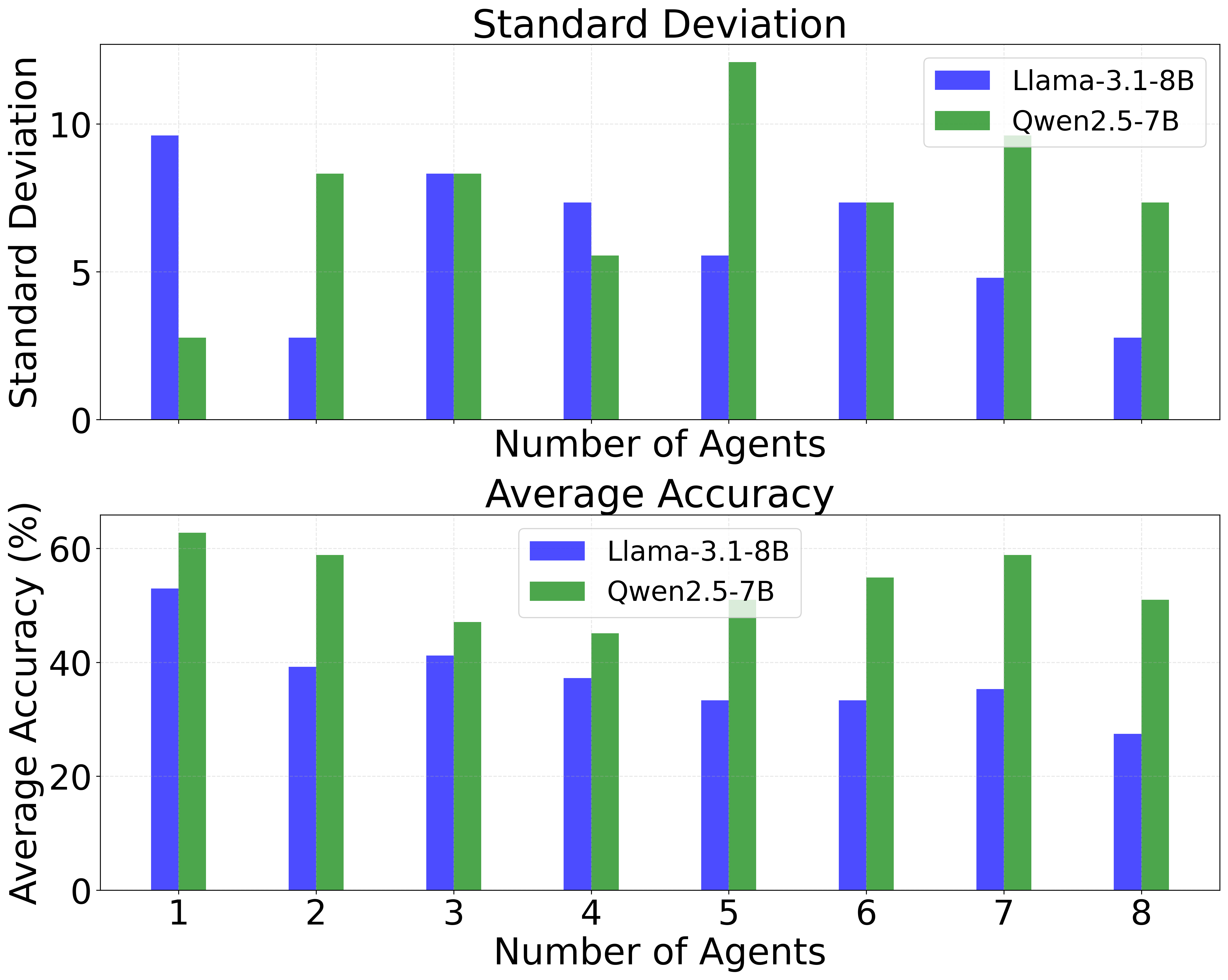}
    \caption{Model Type Impact on College Physics for Models with Small Parameters}
    \label{fig:model_type4_app}
\end{figure}

Our cross-model comparison reveals that scaling behavior is strongly mediated by model architectural differences. 
The Qwen2.5 series demonstrated consistently better adaptation to multi-agent collaboration, achieving earlier performance peaks (optimal at 2 agents for Qwen2.5-72B versus 4 agents for Llama-3.1-70B) and maintaining higher accuracy levels despite agent scaling. 
This advantage manifested through multiple dimensions: Qwen2.5 models maintained superior performance retention (86.27\% peak versus 60.78\% for Llama-3.1), exhibited more stable output patterns (lower standard deviations across agent counts), and demonstrated better resilience to performance degradation at higher agent counts. 
The architectural advantages appear to stem from better complex instruction understanding and iterative reasoning capabilities, enabling more effective information processing in multi-round agent interactions. This finding emphasizes that model selection cannot be separated from multi-agent system design, as inherent architectural characteristics fundamentally shape collaboration dynamics and scaling potential.

\subsection{Agent Configuration Ablation Study}
\label{app:ablation}
\begin{table*}[t]
\centering
\footnotesize
\begin{tabular}{lcccccc}
\toprule
& \multicolumn{3}{c}{\textbf{Accuracy Difference (ALL - Ablation)}} & \multicolumn{3}{c}{\textbf{Std Dev Difference (ALL - Ablation)}} \\
\cmidrule(lr){2-4} \cmidrule(lr){5-7}
\textbf{Subject-Model} & \textbf{Belief} & \textbf{Personality} & \textbf{Expertise} & \textbf{Belief} & \textbf{Personality} & \textbf{Expertise} \\
\midrule
\textbf{Abstract Algebra} & & & & & & \\
\quad Llama3.1 & +9.31 $\uparrow\uparrow$ & +10.78 $\uparrow\uparrow$ & +12.50 $\uparrow\uparrow$ & +2.95 $\uparrow$ & -1.97 $\downarrow$ & +1.27 $\uparrow$ \\
\quad Qwen2.5 & -13.73 $\downarrow\downarrow$ & +0.98 $\rightarrow$ & -4.17 $\downarrow$ & +3.68 $\uparrow$ & -1.99 $\downarrow$ & +3.27 $\uparrow$ \\
\addlinespace

\textbf{College CS} & & & & & & \\
\quad Llama3.1 & +2.57 $\uparrow$ & +18.75 $\uparrow\uparrow$ & +4.84 $\uparrow$ & -3.36 $\downarrow$ & -2.20 $\downarrow$ & -0.33 $\downarrow$ \\
\quad Qwen2.5 & -5.16 $\downarrow$ & -1.24 $\downarrow$ & +0.40 $\rightarrow$ & +2.95 $\uparrow$ & +0.50 $\uparrow$ & -8.16 $\downarrow\downarrow$ \\
\addlinespace

\textbf{College Physics} & & & & & & \\
\quad Llama3.1 & +7.81 $\uparrow$ & +16.64 $\uparrow\uparrow$ & +4.16 $\uparrow$ & +0.74 $\uparrow$ & +1.90 $\uparrow$ & -1.95 $\downarrow$ \\
\quad Qwen2.5 & -10.29 $\downarrow$ & -0.98 $\downarrow$ & -4.66 $\downarrow$ & +2.04 $\uparrow$ & +1.35 $\uparrow$ & +1.57 $\uparrow$ \\
\addlinespace

\textbf{Formal Logic} & & & & & & \\
\quad Llama3.1 & +0.49 $\rightarrow$ & +8.83 $\uparrow$ & -2.82 $\downarrow$ & +7.63 $\uparrow\uparrow$ & +3.58 $\uparrow$ & -2.44 $\downarrow$ \\
\quad Qwen2.5 & -6.37 $\downarrow$ & +1.47 $\uparrow$ & -3.92 $\downarrow$ & +2.26 $\uparrow$ & +0.94 $\uparrow$ & -0.40 $\downarrow$ \\
\addlinespace

\textbf{Global Facts} & & & & & & \\
\quad Llama3.1 & +3.43 $\uparrow$ & +11.27 $\uparrow\uparrow$ & +8.43 $\uparrow$ & +0.34 $\rightarrow$ & +3.61 $\uparrow$ & -1.00 $\downarrow$ \\
\quad Qwen2.5 & +1.96 $\uparrow$ & +4.41 $\uparrow$ & +1.72 $\uparrow$ & +14.95 $\uparrow\uparrow$ & +9.76 $\uparrow\uparrow$ & -4.15 $\downarrow\downarrow$ \\
\addlinespace

\textbf{Philosophy} & & & & & & \\
\quad Llama3.1 & -21.57 $\downarrow\downarrow$ & +15.20 $\uparrow\uparrow$ & -10.09 $\downarrow$ & +4.59 $\uparrow$ & +0.05 $\rightarrow$ & +4.99 $\uparrow$ \\
\quad Qwen2.5 & -23.79 $\downarrow\downarrow$ & +1.22 $\uparrow$ & -11.86 $\downarrow$ & +1.77 $\uparrow$ & -3.34 $\downarrow$ & +0.63 $\uparrow$ \\
\bottomrule
\end{tabular}
\caption{
\footnotesize
Performance Impact of Different Agent Attributes (ALL Mode vs Ablation Modes).
Accuracy Difference = ALL Mode - Ablation Mode; Std Dev Difference = ALL Mode - Ablation Mode. Positive accuracy difference indicates the attribute improves performance. Positive std dev difference indicates the attribute increases instability. \\
Symbols: $\uparrow\uparrow$: significant positive effect ($>|5|$ for accuracy, $>|5|$ for std dev), $\uparrow$: positive effect, $\rightarrow$: minimal effect ($<|2|$), $\downarrow$: negative effect, $\downarrow\downarrow$: significant negative effect ($>|5|$ for accuracy, $>|5|$ for std dev).}
\label{tab:attribute_impact}
\vspace{0.2cm}
\end{table*}
\begin{finding}{1}
\textbf{Agent setting modifications preserve scaling trends but induce performance shifts dependent on discipline and model.} 
\end{finding}

Our experiments reveal that the scaling pattern of MAS—where performance initially improves with more agents but declines beyond an optimal point—remains consistent regardless of agent setting modifications. 
However, the removal of attributes leads to measurable performance shifts that are highly dependent on the task discipline and model type. 
For instance, in abstract algebra with Llama-3.1-8B, removing personality attributes reduced accuracy by an average of 13.58\%, while in philosophy with the same model, it caused a drastic accuracy drop of 21.57\%. Conversely, with Qwen2.5-7B, personality removal had minimal impact on accuracy (average +0.98\%) but increased instability. 
Similarly, belief removal exacerbated performance declines in philosophy for both models (up to -23.79\% for Qwen2.5-7B) but improved accuracy in some technical disciplines like college physics for Llama-3.1-8B (+7.81\%). 
This finding underscores that while agent settings do not change the fundamental scaling dynamics, they introduce discipline-specific and model-dependent performance variations, necessitating tailored configurations for optimal results.

\begin{finding}{2}
\textbf{Specific agent attributes have distinct impacts: personality enhances accuracy at the cost of stability, beliefs introduce task-dependent biases which can degrade performance, and expertise yields mixed, model-specific effects.} 
\end{finding}

Detailed analysis of individual attributes reveals distinct impacts on system performance. 
Personality attributes generally boost accuracy—particularly for models like Llama-3.1-8B, where they improved accuracy by an average of 13.58 percentage points across disciplines—by fostering diverse perspectives and critical thinking. 
However, this came at the cost of reduced stability, as personality increased standard deviations by an average of 0.83\% for Llama-3.1-8B and 1.20\% for Qwen2.5-7B, due to increased behavioral variability and conflict. 
Belief attributes, while potentially guiding reasoning in logical tasks (e.g., +9.31\% in abstract algebra for Llama-3.1-8B), often introduced instability and performance degradation in subjective disciplines like philosophy (accuracy drops up to -23.79\% for Qwen2.5-7B), as beliefs exacerbated disagreements among agents. 
Expertise attributes showed model-dependent effects: they benefited Llama-3.1-8B in technical disciplines (e.g., +12.50\% in abstract algebra) but hindered Qwen2.5-7B in abstract tasks (e.g., -4.17\% in philosophy), indicating that expertise can either focus reasoning or constrain flexibility based on model architecture. 
These findings emphasize that agent settings must be carefully tuned, with personality suitable for accuracy-critical tasks, beliefs used cautiously in objective domains, and expertise aligned with model strengths to balance performance and stability.

In summary, agent settings play a crucial role in fine-tuning multi-agent system performance, but their effects are nuanced and context-dependent. System designers should prioritize attribute configuration based on task characteristics and model capabilities, leveraging attributes like personality for accuracy gains in reasoning-intensive tasks while mitigating stability risks through iterative testing and calibration.

\section{Supplementary Material for Comparison}
\label{app:comp}

\subsection{CoT Prompts}
\label{app:cot}
This section details the three Chain-of-Thought (CoT) prompt templates used in our experiments. The placeholders (e.g., \texttt{\{problem.question\}}) are replaced with concrete content during execution.

\textbf{1. Simple CoT}

This template guides the model to produce step-by-step reasoning followed by a structured final answer, suitable for objective questions (e.g., multiple-choice, short-answer).

\vspace{\baselineskip}

\begin{ACLPrompt}
\textbf{System Instruction:}\\
You are \textit{\{self.name\}}. Please solve the following problem in a reasoning manner.\\
\textit{\{message.content\}}

\vspace{1ex}
\textbf{Output Format Requirement:}\\
Please structure your response as (without '[]' in your response):

\textbf{\#\#\# Reasoning}\\
\textit{[Your reasoning process for arriving at the final answer, including any calculations or logical deductions]}

\textbf{\#\#\# Answer}\\
\textit{[Your final answer for the given problem in accordance with the required form, such as \{"A, B, C, or D for a multiple-choice question"\} if \texttt{self.question\_type == 'multiple\_choice'} else "direct answer like '100' for a short-answer question", without any explanation]}
\end{ACLPrompt}

\captionof{figure}{Simple CoT prompt template.}
\label{lst:cot-simple}

\vspace{\baselineskip}

\textbf{2. Coding CoT}

This template instructs the model to analyze a coding problem step-by-step and provide a complete, efficient solution with code.

\vspace{\baselineskip}

\begin{ACLPrompt}
\textbf{System Instruction:}\\
Please solve the following coding problem. Use step-by-step reasoning and provide a complete, efficient code solution.

\textbf{Problem:}\\
\textit{\{problem.question\}}

\textbf{Programming language:} \textit{\{language\}}\\
\textbf{Constraints:}\\
\textit{\{constraints\_str\}}

\vspace{1ex}
\textbf{Your solution should:}\\
\textbullet~ Fully satisfy the problem requirements\\
\textbullet~ Be clear and readable\\
\textbullet~ Be efficient and correct\\
\textbullet~ Include necessary comments

\vspace{1ex}
\textbf{Please organize your response as follows:}\\
\textbf{\#\#\# Reasoning Process}\\
\textit{[Your analysis]}

\textbf{\#\#\# Final Code}\\
\textit{[Complete code solution]}
\end{ACLPrompt}

\captionof{figure}{Coding CoT prompt template.}
\label{lst:cot-coding}

\vspace{\baselineskip}

\textbf{3. Open-ended CoT}

This template prompts the model for a comprehensive and insightful analysis of open-ended questions, guided by specific evaluation criteria.

\vspace{\baselineskip}

\begin{ACLPrompt}
\textbf{System Instruction:}\\
Please analyze the following open-ended question in depth. Use step-by-step reasoning and provide a comprehensive and insightful response.

\textbf{Question:}\\
\textit{\{problem.question\}}

\textbf{Evaluation criteria:}\\
\textit{\{criteria\_str\}}

\vspace{1ex}
\textbf{Your response should:}\\
\textbullet~ Be thorough and insightful\\
\textbullet~ Be well-structured\\
\textbullet~ Demonstrate innovation and practicality\\
\textbullet~ Align with the evaluation criteria

\vspace{1ex}
\textbf{Please organize your response as follows:}\\
\textbf{\#\#\# Reasoning Process}\\
\textit{[Your analysis]}

\textbf{\#\#\# Final Answer}\\
\textit{[A direct and comprehensive answer to the question]}
\end{ACLPrompt}

\captionof{figure}{Open-ended CoT prompt template.}
\label{lst:cot-open}

\vspace{\baselineskip}

\subsection{Evaluation Protocol for Generation Tasks}
\label{app:gen_eval}
The dataset comprises a custom collection of 10 coding problems and 15 open-ended questions. 
The coding problems involve 5 algorithmic problems and 5 complex software development projects that go beyond simple algorithmic implementation, while the open-ended questions pertain to product ideation and solutions for societal issues. 
The evaluation of results is entrusted to a dedicated assessment agent. This agent scores the Chain-of-Thought reasoning and SIMAS responses against predefined criteria for each problem. Based on the scores, it determines which approach prevails on a given question and provides a justification for its judgment. 
Figure~\ref{fig:code_comparison_app} shows the results of experiments tested on the 5 complex software development projects, and Figure~\ref{fig:open_ended_comparison_app} shows that on the 15 open-ended questions.
Examples from the dataset are provided in Table~\ref{tab:code_problems} and ~\ref{tab:open_ended_problems}.
The prompt is shown in List~\ref{lst:evaluator}:
\vspace{\baselineskip}

\begin{ACLPrompt}
\textbf{System Role:}\\
As a professional evaluation expert, please compare two AI-generated responses to the same problem.

\textbf{Problem:}\\
\textit{\{problem\}}

\textbf{Response A} (generated by \textit{\{method\_a\}}):\\
\textit{\{answer\_a\}}

\textbf{Response B} (generated by \textit{\{method\_b\}}):\\
\textit{\{answer\_b\}}

\textbf{Evaluation criteria} (each criterion scored 1-10):\\
\textit{\{criteria\_str\}}

\vspace{1ex}
\textbf{Instructions:}\\
1. Evaluate each response on every criterion separately.\\
2. Compare the strengths and weaknesses of both responses.\\
3. Indicate which response is better and provide detailed reasoning.

\vspace{1ex}
\textbf{Return the results in the following JSON format:}\\
{\ttfamily \small
\{ \\
\hspace*{4mm}"response\_a\_scores": \{ \\
\hspace*{8mm}"total": X, \\
\hspace*{8mm}"criteria": \{ \\
\hspace*{12mm}"criterion1": \{"score": X, "reason": "..."\}, \\
\hspace*{12mm}... \\
\hspace*{8mm}\} \\
\hspace*{4mm}\}, \\
\hspace*{4mm}"response\_b\_scores": \{ \\
\hspace*{8mm}"total": X, \\
\hspace*{8mm}"criteria": \{ \\
\hspace*{12mm}"criterion1": \{"score": X, "reason": "..."\}, \\
\hspace*{12mm}... \\
\hspace*{8mm}\} \\
\hspace*{4mm}\}, \\
\hspace*{4mm}"comparison": \{ \\
\hspace*{8mm}"winner": "A" or "B" or "tie", \\
\hspace*{8mm}"reason": "...", \\
\hspace*{8mm}"key\_differences": ["...", "..."], \\
\hspace*{8mm}"potential\_synergy": "..." \\
\hspace*{4mm}\} \\
\}
}

\vspace{1ex}
\textbf{Constraints:}\\
Please ensure the evaluation is fair, objective, and based solely on the content of the responses, not the generation method. Ensure the response contains only valid JSON format with no additional text or symbols.
\end{ACLPrompt}

\captionof{figure}{Prompt template for AI response evaluation.}
\label{lst:evaluator}

\vspace{\baselineskip}

\subsection{Full Comaprison Results}
\begin{figure*}[h]
    \centering
    \begin{subfigure}[b]{\columnwidth}
        \centering
        \includegraphics[width=\textwidth]{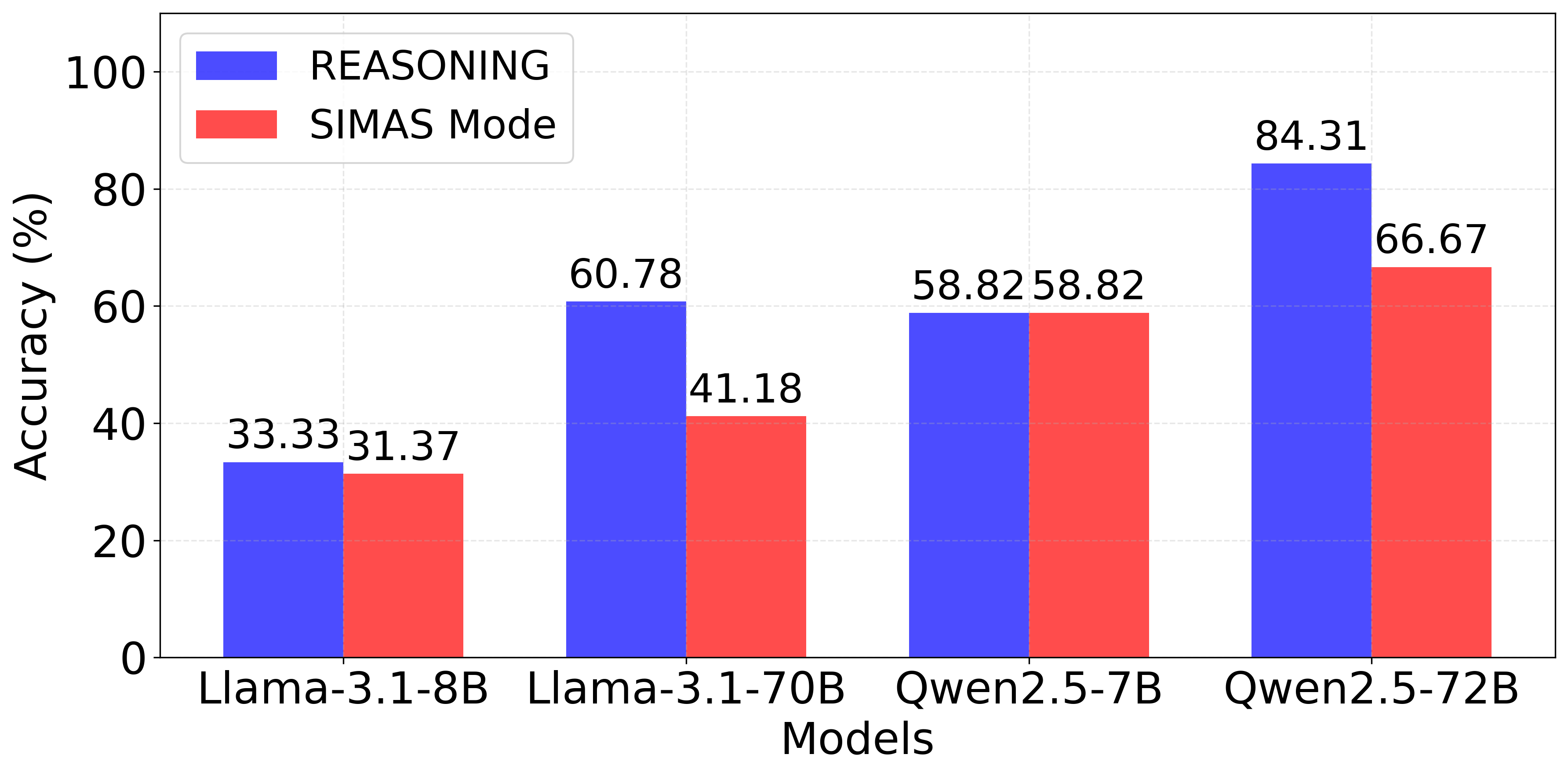}
        \caption{CoT vs. SIMAS on Abstract Algebra (Llama-3.1-70B). CoT consistently outperforms all multi-agent configurations.}
        \label{fig:cot_vs_simas_algebra_app}
    \end{subfigure}
    \hfill
    \begin{subfigure}[b]{\columnwidth}
        \centering
        \includegraphics[width=\textwidth]
        {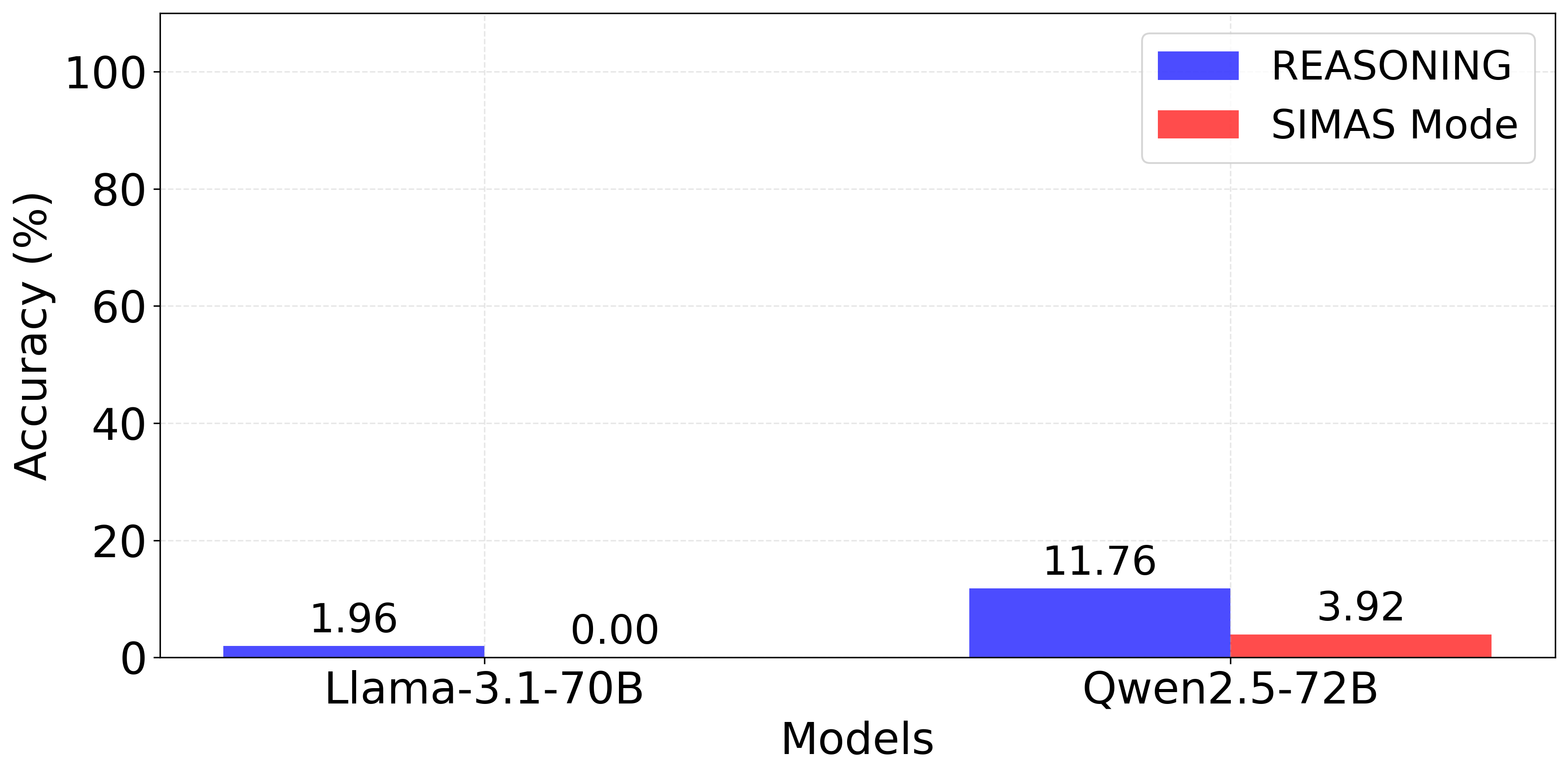}
        \caption{CoT vs. SIMAS on AIME 2025. SIMAS fails catastrophically, highlighting its inadequacy for complex, multi-step reasoning.}
        \label{fig:cot_vs_simas_aime_app}
    \end{subfigure}
    
    \vspace{0.3cm} 
    
    \begin{subfigure}[b]{\columnwidth}
        \centering
        \includegraphics[width=\textwidth]{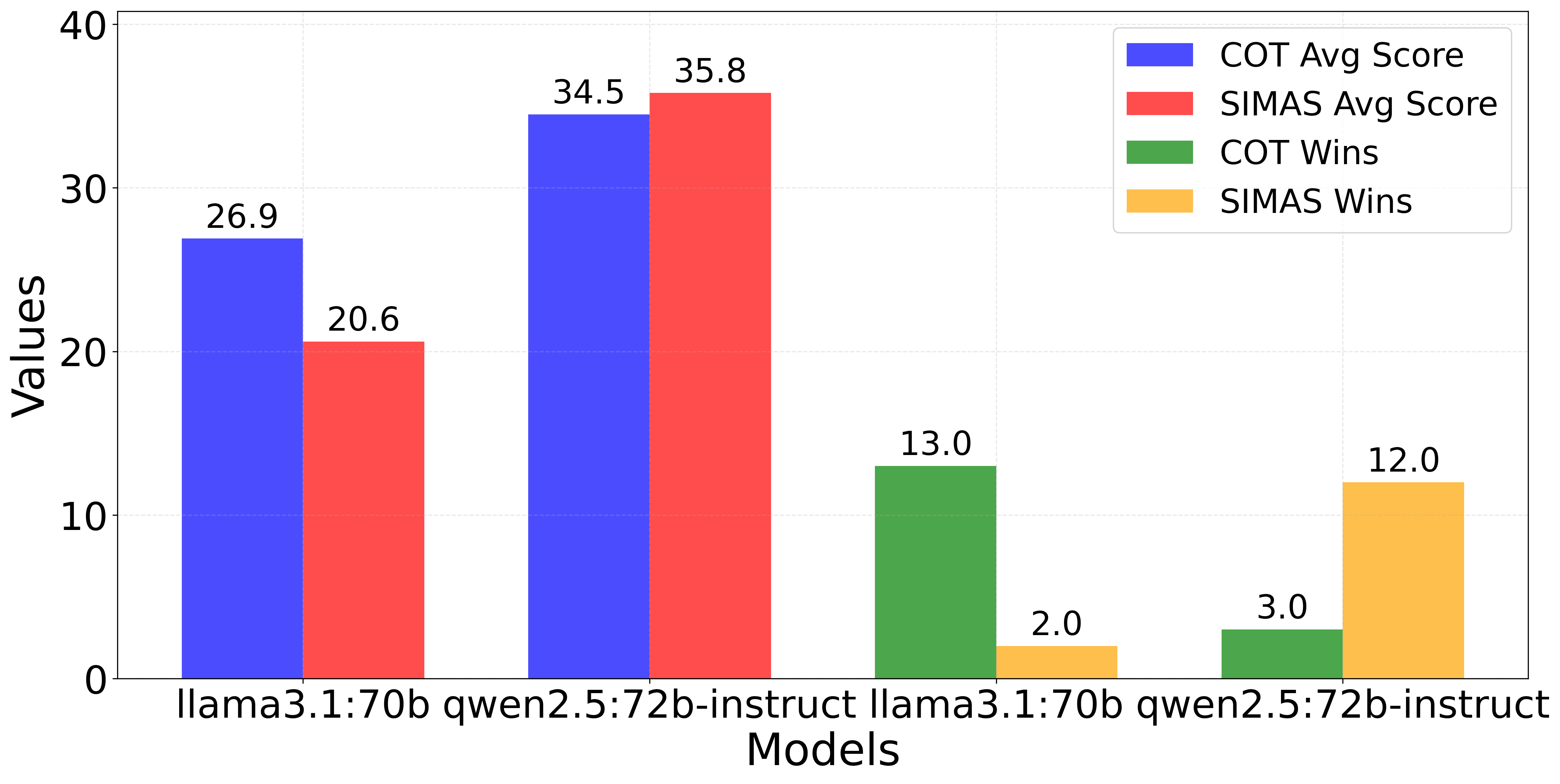}
        \caption{Win-rate and average score comparison on open-ended tasks. Results varies depending on model type.}
        \label{fig:open_ended_comparison_app}
    \end{subfigure}
    \hfill
    \begin{subfigure}[b]{\columnwidth}
        \centering
        \includegraphics[width=\textwidth]{figures/code_comparison_combined.png}
        \caption{Win-rate comparison on coding tasks. SIMAS shows a model-dependent, marginal advantage, highlighting inconsistent benefits of simple collaboration.}
        \label{fig:code_comparison_app}
    \end{subfigure}
\end{figure*}
On reasoning-intensive domains such as abstract algebra and advanced mathematics, where our SIMAS framework lacks corrective mechanisms, single-agent reasoning demonstrates clear superiority. As shown in Figure~\ref{fig:cot_vs_simas_algebra_app}, for abstract algebra, the CoT baseline consistently outperformed the best-performing SIMAS configuration. 
This gap escalates into a decisive failure for SIMAS on AIME 2025 (Figure~\ref{fig:cot_vs_simas_aime_app}), where it fragments logical chains without a means to reintegrate them. 
The results on open-ended creative tasks (Figure~\ref{fig:open_ended_comparison_app}) are similarly unequivocal: single-agent reasoning vastly outperformed SIMAS, as narrative coherence is shattered by unmoderated multi-agent turn-taking.

The coding task (Figure~\ref{fig:code_comparison_app}) and open-ended task (Figure~\ref{fig:open_ended_comparison_app}) domains present a more contested picture for SIMAS, with marginal gains for Llama-3.1-70B on coding and Qwen-2.5-72B on open-ended tasks. 
The dataset comprises a custom collection of 5 coding problems and 15 open-ended questions. The coding problems involve complex software development projects that go beyond simple algorithmic implementation, while the open-ended questions pertain to product ideation and solutions for societal issues. The evaluation of results is entrusted to a dedicated assessment agent. This agent scores the Chain-of-Thought reasoning and SIMAS responses against predefined criteria for each problem. Based on the scores, it determines which approach prevails on a given question and provides a justification for its judgment. Examples from the dataset are provided in the Appendix.

The results suggest that even a simple sequential architecture can sometimes add value for tasks benefiting from multi-perspective validation. However, the qualitative analysis of agent conversations reveals the inefficiency of this gain. 
A case study of a discussion on the ``Two Sum'' problem showed early convergence on the optimal solution, followed by rounds of redundant meta-discussion and unproductive speculation on inferior alternatives. The final output was functionally identical to the initial proposal. This pattern illustrates that without an architecture designed to synthesize, critique, and refine efficiently, multi-agent dialogue often devolves into \textbf{low-value redundancy}, consuming context window capacity without generating synergistic insight. 
The marginal win comes at a disproportionately high computational cost and is not reliably generalizable across models, as shown by Qwen2.5-72B's preference for CoT on coding tasks.

\subsection{Case Study: Dialogue for Solving Mathematical Problem}
\label{app:dialogue_case_reasoning}
To illustrate the fragmentation of coherent thought in multi-agent reasoning, we present an excerpt from a SIMAS discussion on a polynomial root-finding problem in a finite field. The task requires finding all zeros of \(x^5 + 3x^3 + x^2 + 2x\) in \(\mathbb{Z}_5\), with four multiple-choice options: A) 0, B) 1, C) 0,1, D) 0,4. The complete dialogue history is exhibited in Table~\ref{tab:dialogue_history_reasoning}. The correct answer is D) 0,4, as verified by evaluating the polynomial at each element of \(\mathbb{Z}_5\): \(x=0\) yields 0, \(x=4\) yields \(4^5+3\cdot4^3+4^2+2\cdot4 \equiv 4+2+1+3 \equiv 0 \pmod{5}\). However, the multi-agent discussion resulted in an incorrect conclusion (A) 0), demonstrating a critical failure in reasoning continuity.

\textbf{Summary of Dialogue Dynamics:}
The conversation spanned three rounds among three agents (Bob, Charlie, Diana). In Round 1, Charlie correctly initiated the process of checking each element of \(\mathbb{Z}_5\) and verified \(x=0\) as a root. Diana then checked \(x=1\) and correctly concluded it was not a root. In Round 2, Diana attempted to check \(x=2\) but made a critical error: she evaluated the expression \(2^3 + 2 \cdot 2 + 1 = 13 \equiv 3 \pmod{5}\), which does not correspond to the original polynomial \(x^5 + 3x^3 + x^2 + 2x\). This mistake went unnoticed by both Bob and Charlie, who accepted the result without verification. Moreover, the agents never examined \(x=3\) or \(x=4\). In Round 3, Bob summarized that the only root was \(x=0\), and he further compounded the error by referring to the polynomial as \(x^3 + 2x + 1\)—a completely different polynomial. Charlie and Diana then concluded the discussion with expressions of gratitude and regret, respectively, without any correction or completion of the required checks.

\textbf{Analysis of Fragmentation:}
This transcript exemplifies the core failure mode of fragmented reasoning:
1. \textbf{Loss of Logical Continuity:} A coherent reasoning chain was established in Round 1, but it was broken in Round 2 when Diana introduced an erroneous computation and failed to check all elements. Instead of grounding the discussion in the original problem and systematically verifying each candidate, subsequent agents built upon the incomplete and flawed fragments, leading to a solution that missed the root \(x=4\).
2. \textbf{Absence of Synthesis and Correction:} No agent synthesized the partial results into a unified, correct line of reasoning. The verification-oriented persona evident in the earlier algebra example was absent here; the agents lacked a mechanism to detect the error in Diana's evaluation or to recognize that \(x=3\) and \(x=4\) remained unchecked. Consequently, the discussion drifted further from the correct solution path, culminating in an incorrect final answer.
3. \textbf{Divergent Drift from Task Constraints:} The original problem specified the polynomial \(x^5 + 3x^3 + x^2 + 2x\), yet by the final round, Bob was discussing \(x^3 + 2x + 1\). This drift occurred without any acknowledgment or correction, illustrating how sequential turn-taking without a grounding mechanism can lead to unrecognized solution divergence and outright errors.
4. \textbf{Incomplete Verification:} The agents failed to exhaustively test all elements of \(\mathbb{Z}_5\) (0,1,2,3,4), stopping after a mistaken evaluation of \(x=2\). This incomplete verification, combined with the lack of cross-checking, directly caused the omission of the valid root \(x=4\).

This case underscores that in multi-step reasoning tasks, the absence of a dedicated synthesis or error-correction mechanism causes the collective output to become disjointed and error-prone. The final answer (A) 0) is incorrect for the original problem, highlighting that even when the reasoning process appears collaborative, the fragmentation of coherent thought can lead to reliable-sounding but fundamentally flawed conclusions. The agents' failure to maintain a consistent logical thread and to correct accumulating errors demonstrates the need for structured interaction protocols that enforce systematic verification and synthesis.

\subsection{Case Study: Dialogue for Solving Coding Problem}
\label{app:dialogue_case}
To illustrate the inefficiencies that can arise in unstructured multi-agent dialogue, we present an excerpt from a SIMAS discussion on the classic ``Two Sum'' coding problem. The task requires finding two indices in an array whose values sum to a target, with constraints of $O(n)$ time and $O(n)$ space complexity. The whole dialogue history is exhibited in Table~\ref{tab:dialogue_history}.

\textbf{Summary of Dialogue Dynamics:}
The conversation spanned three rounds among three agents (Nova, Agent\_2, Riven). In the first round, Agent\_2 correctly identified the core solution strategy (using a hash table/dictionary). In the second round, Nova promptly provided a complete, correct implementation of the hash table solution, meeting all specified constraints. However, instead of converging, the discussion continued into a third round where Riven raised concerns about edge cases (e.g., empty input, duplicate numbers) that are explicitly ruled out by the problem assumptions (``each input has exactly one solution''). Nova then added an unnecessary empty-list check, which did not improve the solution's correctness under the given constraints. The final answer was substantively identical to the initial implementation, albeit with minor cosmetic changes and redundant commentary.

\textbf{Analysis of Inefficiencies:}
This transcript exemplifies several systemic issues:
1. \textbf{Low-Value Redundancy:} The core algorithm was identified and implemented early, yet the conversation continued for multiple rounds without introducing new insights or improvements.
2. \textbf{Misplaced Critique:} Agents spent time discussing edge cases that were irrelevant under the problem's explicit assumptions, illustrating a failure to ground discussion in the given task constraints.
3. \textbf{Inefficient Use of Context:} The progressive history accumulation led to lengthy meta-discourse and repetitions (e.g., Agent\_2 repeatedly endorsing the hash table approach) that consumed token budget without advancing the solution.

This case underscores that without architectural mechanisms to synthesize information, critique productively, and terminate upon convergence, multi-agent dialogues can devolve into ceremonious discussion rather than efficient problem-solving. The marginal refinement observed came at a disproportionately high computational cost, highlighting the need for structured interaction protocols beyond simple sequential turn-taking.

\begin{table*}[t]
\centering
\begin{tabularx}{\textwidth}{llXl}
\toprule
\textbf{ID} & \textbf{Category} & \textbf{Question (English)} & \textbf{Evaluation Criteria} \\
\midrule
\multirow{2}{*}{1} & \multirow{2}{*}{algorithm} & 
Implement a function that takes an integer array \texttt{nums} and a target value \texttt{target}, and returns the indices of two numbers in the array that sum up to \texttt{target}. 

\textbf{Requirements:}
1. Assume each input has exactly one solution.
2. Cannot use the same element twice.
3. Time complexity should be as low as possible.

\textbf{Example:}
\texttt{nums = [2, 7, 11, 15], target = 9}
\texttt{Return: [0, 1]} &
\makecell[l]{Correctness\\ Time complexity\\ Code simplicity\\ Boundary case handling} \\
\midrule
4 & string\_algorithm & 
Implement a simple regular expression matcher. Support the following two wildcards:
- '.' matches any single character
- '*' matches zero or more of the preceding element

Implement function: \texttt{def is\_match(s: str, p: str) -> bool:}
where \texttt{s} is the string to match and \texttt{p} is the pattern. &
\makecell[l]{Algorithm correctness\\ Boundary case coverage\\ Time complexity\\ Code readability} \\
\midrule
8 & iot\_backend & 
Design a backend system for monitoring IoT devices and an intelligent rule engine. The system should receive and process data streams from simulated or real IoT sensors (e.g., temperature, humidity, motion). It must implement data persistence, real-time threshold alerts, and provide a rule engine that allows users to define complex interaction rules through a graphical interface or DSL (e.g., 'When living room temperature > 30°C and motion detected, turn on AC and send notification'). &
\makecell[l]{Data stream processing capability \\ and stability\\ Flexibility and expressiveness of \\ rule engine design\\ System security considerations\\ (device authentication, data privacy)\\ API design and frontend management \\ interface usability\\ System scalability and fault tolerance} \\
\bottomrule
\end{tabularx}
\caption{Sample Programming Problems with Evaluation Criteria}
\label{tab:code_problems}
\end{table*}

\begin{table*}[t]
\centering
\begin{tabularx}{\textwidth}{llXl}
\toprule
\textbf{ID} & \textbf{Category} & \textbf{Question (English)} & \textbf{Evaluation Criteria} \\
\midrule
1 & urban\_planning & 
Design an innovative urban transportation solution to address peak-hour congestion. Consider sustainability, cost-effectiveness, and user experience. &
\makecell[l]{Innovativeness\\ Feasibility\\ Cost-effectiveness\\ Sustainability\\ User experience} \\
\midrule
6 & sustainability & 
Design a community-shared resource platform under a circular economy model, aiming to maximize resource utilization and reduce waste. Explain its operational mechanism, incentive measures, and community governance structure. &
\makecell[l]{Systems thinking\\ Model innovativeness\\ Community participation\\ Environmental benefit quantification} \\
\midrule
10 & technology\_society & 
Discuss the potential impacts of artificial intelligence (e.g., machine translation, content generation) on global minority languages and cultural heritage preservation, and propose coping strategies, while considering the protection of cultural diversity. &
\makecell[l]{Dialectical insight\\ Cultural sensitivity\\ Ethical considerations of \\ technical solutions\\ Multi-level strategy formulation} \\
\bottomrule
\end{tabularx}
\caption{Sample Open-ended Problems with Evaluation Criteria}
\label{tab:open_ended_problems}
\end{table*}

```latex
\begin{table*}[ht]
\centering
\small
\newcolumntype{Y}{>{\raggedright\arraybackslash}X}
\begin{tabularx}{\textwidth}{|c|c|Y|}
\hline
\textbf{Round} & \textbf{Agent} & \textbf{Dialogue Content} \\
\hline
\multirow{3}{*}{1} & Bob & Hello everyone! Let's discuss the following problem:\newline\newline Find all zeros in the indicated finite field of the given polynomial with coefficients in that field. \(x^5 + 3x^3 + x^2 + 2x\) in \(\mathbb{Z}_5\)\newline A) 0\newline B) 1\newline C) 0,1\newline D) 0,4 \\
\cline{2-3}
 & Charlie & Hi Bob! Let's dive into this problem. To find the zeros of the polynomial \(x^5 + 3x^3 + x^2 + 2x\) in \(\mathbb{Z}_5\), we need to check each element of \(\mathbb{Z}_5\) (i.e., 0, 1, 2, 3, 4) and see which ones make the polynomial equal to zero. Let's start with \(x = 0\). \\
\cline{2-3}
 & Diana & Hi Bob and Charlie! I'm eager to contribute. Let's continue with checking the other elements in \(\mathbb{Z}_5\). For \(x = 1\), we substitute into the polynomial: \(1^5 + 3(1^3) + 1^2 + 2(1) = 1 + 3 + 1 + 2 = 7 \equiv 2 \pmod{5}\), which is not zero. So, 1 is not a root. \\
\hline
\multirow{3}{*}{2} & Bob & Great work, everyone! We've confirmed that \(x = 0\) is a root, and \(x = 1\) is not a root. Let's continue checking the remaining elements in \(\mathbb{Z}_5\): \(x = 2\), \(x = 3\), and \(x = 4\). Diana, you can start with \(x = 2\). \\
\cline{2-3}
 & Charlie & Sounds good, Bob! Let's keep going. Diana, I'm looking forward to seeing your results for \(x = 2\). \\
\cline{2-3}
 & Diana & I'll get right on it. Let's see what happens when \(x = 2\). Plugging in the value, we have \(2^3 + 2 \cdot 2 + 1 = 8 + 4 + 1 = 13\), which is equivalent to \(3\) modulo 5. Therefore, \(x = 2\) is not a root of the polynomial in \(\mathbb{Z}_5\). \\
\hline
\multirow{3}{*}{3} & Bob & Great job, everyone! We've now determined that \(x = 0\) is the only root of the polynomial \(x^3 + 2x + 1\) in \(\mathbb{Z}_5\). The other elements \(x = 1\), \(x = 2\), \(x = 3\), and \(x = 4\) are not roots. This thorough analysis has been very productive, and I appreciate the effort and accuracy from both Charlie and Diana. \\
\cline{2-3}
 & Charlie & Thank you, Bob! It's great to hear that our collaborative effort has been successful. I learned a lot from this process, and I'm glad we were able to determine the roots of the polynomial accurately. Looking forward to more such discussions! \\
\cline{2-3}
 & Diana & Thank you both for your efforts and the positive outcome. I regret not being able to contribute to the discussion and will make sure to participate more actively in future conversations. \\
\hline
\end{tabularx}
\caption{Dialogue History Across Three Rounds}
\label{tab:dialogue_history_reasoning}
\end{table*}
```

\begin{table*}[ht]
\centering
\small
\begin{tabularx}{\textwidth}{|c|c|Y|}
\hline
\textbf{Round} & \textbf{Agent} & \textbf{Dialogue Content} \\
\hline
\multirow{3}{*}{1} & Nova & Let's discuss this coding problem: Implement a function that accepts an integer array \texttt{nums} and a target value \texttt{target}, and returns the indices of two numbers in the array that sum to \texttt{target}. Requirements: 1. Assume each input has exactly one solution. 2. Cannot use the same element twice. 3. Time complexity should be as low as possible. Example: \texttt{nums = [2, 7, 11, 15], target = 9}, return \texttt{[0, 1]}. Programming language: Python. Constraints: \texttt{['time complexity O(n)', 'space complexity O(n)']}. Please share your solution. \\
\cline{2-3}
 & Agent\_2 & Nova's problem statement seems well-defined and clear. I agree that breaking down complex code into manageable parts for analysis is crucial in solving this problem efficiently. Since the goal is to find two numbers in an array that add up to a target value with a low time complexity, I think we can explore using hash tables or dictionaries to store the numbers we've seen so far and their indices. This would allow us to check if the complement of the current number (i.e., the number that adds up to the target) has been seen before in constant time. What are your thoughts on this approach, Nova? Have you considered using any data structures to improve the efficiency of our solution? \\
\cline{2-3}
 & Riven & Alright, let's get down to business. Agent\_2, your idea of using hash tables or dictionaries sounds good in theory, but how do you plan to implement it? What specific data structure would you use, and how would you handle edge cases? Nova, I'd like to see some actual code from you as well. A problem statement is just the beginning -- let's see how you intend to solve this problem efficiently. Let's not waste time on abstract discussions; show me the code, and then we can talk about optimization and critique. \\
\hline
\multirow{3}{*}{2} & Nova & I appreciate Agent\_2's feedback and suggestions on using hash tables or dictionaries to improve the solution's efficiency. Riven, I understand your request for concrete code, so here's my attempt at a Python solution: \texttt{def two\_sum(nums, target): num\_dict = \{\} for i, num in enumerate(nums): complement = target - num if complement in num\_dict: return [num\_dict[complement], i] num\_dict[num] = i return []}. This solution uses a dictionary to store the numbers we've seen so far and their indices. We iterate through the input list, calculate the complement of each number with respect to the target, and check if it exists in the dictionary. If it does, we return the indices of the two numbers that add up to the target. Please let me know if this solution meets your requirements or if there's room for improvement. \\
\cline{2-3}
 & Agent\_2 & Nova's solution using a dictionary to store numbers and their indices is a good implementation of the hash table concept we discussed earlier. I'm glad to see that my suggestions were helpful in improving the efficiency of the solution. Riven, since Nova has provided a concrete code example, would you like to review it for any potential flaws or areas for optimization? Your expertise in code critique and optimization would be valuable in ensuring the solution is robust and reliable. \\
\cline{2-3}
 & Riven & Finally, some concrete code to sink my teeth into. Nova, your solution looks decent, but I'm not here to coddle you. Let's dive into the nitty-gritty. Firstly, using a dictionary is a good call, as it allows for efficient lookups. However, have you considered the scenario where the input list contains duplicate numbers? Your current implementation would only store the index of the last occurrence of each number. This might not be the desired behavior in certain cases. Additionally, I'd like to see some error handling. What if the input list is empty or None? What if the target is not a valid number? You should add some checks to ensure your function behaves correctly under these conditions. Lastly, while your solution has a time complexity of O(n), where n is the length of the input list, I'm not convinced it's the most efficient approach. Have you considered using a different data structure or algorithm? Let me hear your thoughts on these concerns, Nova. Agent\_2, feel free to chime in as well. \\
\hline
\multirow{3}{*}{3} & Nova & I appreciate the thorough review of my solution by Riven and Agent\_2's encouragement. To address Riven's concerns, I agree that handling duplicate numbers and edge cases such as an empty or None input list is crucial. Here's an updated version of the code: \texttt{def two\_sum(nums, target): if not nums: return [] num\_dict = \{\} for i, num in enumerate(nums): complement = target - num if complement in num\_dict: return [num\_dict[complement], i] num\_dict[num] = i return []}. Regarding the use of a different data structure or algorithm, I'm open to exploring alternative approaches that could improve efficiency. However, based on my analysis, the dictionary-based solution seems to be a reasonable trade-off between time complexity and readability. I'd like to thank both Agent\_2 and Riven for their valuable feedback, which has helped me refine my solution. \\
\cline{2-3}
 & Agent\_2 & I'm pleased to see that our collaborative effort has led to a refined solution. Nova's updated code effectively addresses edge cases and maintains readability while achieving optimal time complexity. As we conclude this discussion, I'd like to reiterate the importance of systematic analysis and logical thinking in problem-solving. By breaking down complex problems into manageable parts and applying mathematical rigor, we can consistently arrive at efficient solutions. Thank you, Nova and Riven, for your contributions to this conversation. \\
\cline{2-3}
 & Riven & Nova's updated solution addresses some of the concerns I had, but I still think there's room for improvement in terms of efficiency. Using a dictionary-based approach might be reasonable, but have you considered exploring other data structures that could potentially lead to even faster solutions? Agent\_2, while your emphasis on systematic analysis is commendable, I believe our discussion could have benefited from more aggressive critiquing of the code. Nevertheless, I appreciate the collaborative effort and acknowledge that it has led to a refined solution. \\
\hline
\end{tabularx}
\caption{Dialogue History Across Three Rounds for Coding.}
\label{tab:dialogue_history}
\end{table*}
\end{CJK}
\end{document}